\documentclass[lettersize,journal]{IEEEtran}
\usepackage{amsmath,amsfonts}
\usepackage{array}
\usepackage[caption=false,font=normalsize,labelfont=sf,textfont=sf]{subfig}
\usepackage{textcomp}
\usepackage{stfloats}
\usepackage{url}
\usepackage{bm}      % For bold math symbols
\usepackage{booktabs} % For better-looking tables
\usepackage{algorithm} 
\usepackage[table]{xcolor}
\usepackage{siunitx}
\usepackage{multirow}
\usepackage{makecell}  % 核心换行包
\usepackage{verbatim}
\usepackage{romannum}

\usepackage{graphicx}
\usepackage{algpseudocode}
\usepackage{amsmath,amssymb}
\graphicspath{
	{./}                % 当前目录
	{Figure/}           % 当前目录下的Figure子文件夹
}

\def\BibTeX{{\rm B\kern-.05em{\sc i\kern-.025em b}\kern-.08em
    T\kern-.1667em\lower.7ex\hbox{E}\kern-.125emX}}
\usepackage{balance}

\begin{document}
\pagenumbering{arabic}
\title{Delay-Doppler Domain Channel Measurements and Modeling in High-Speed Railways}
\author{Hao Zhou, \emph{Graduate Student Member, IEEE}, Yiyan Ma, \emph{Member, IEEE}, Dan Fei, \emph{Member, IEEE}, \\ Weirong Liu, \emph{Graduate Student Member, IEEE}, Zhengyu Zhang, \emph{Graduate Student Member, IEEE}, \\ Mi Yang, \emph{Member, IEEE}, Guoyu Ma, \emph{Member, IEEE}, Yunlong Lu, \emph{Member, IEEE}, Ruisi He, \emph{Senior Member, IEEE,} Guoyu Wang, Cheng Li, Zhaohui Song, \emph{Senior Member, IEEE,} Bo Ai, \emph{Fellow, IEEE}
\thanks{This work is supported by the the National Natural Science Foundation of China under Grant U25A20279, 62341127, 62221001 and 62501041; the State Key Laboratory of Advanced Rail Autonomous Operation Research Funds (Contract No.RAO2023ZZ004); the Fundamental Research Funds for the Natural Science Foundation of Jiangsu Province, Major Project under Grant BK20212002; the Fundamental Research Funds for the Central Universities 2025XKBH012; the China Postdoctoral Science Foundation (GZB20250807, 2025M783497); the open research fund of National Mobile Communications Research Laboratory, Southeast University (No. 2026D04). An earlier version of this article was presented at the 2025 IEEE/CIC International Conference on Communications in China (ICCC Workshops), China [DOI: 10.1109/ICCCWorkshops67136.2025.11148166]. \emph{(Corresponding authors: Yiyan Ma.)}

H. Zhou, Y. Ma, D. Fei, W. Liu, Z. Zhang, M. Yang, G. Ma, Y. Lu, R. He, G. Wang, C. Li and B. Ai are with the State Key Laboratory of Advanced Rail Autonomous	Operation and the School of Electronic and Information Engineering, Beijing Jiaotong University, Beijing 100044, China. (email: hao.zhou@bjtu.edu.cn; mayiyan@bjtu.edu.cn; dfei@bjtu.edu.cn; 22110040@bjtu.edu.cn; 21111040@bjtu.edu.cn; myang@bjtu.edu.cn; magy@bjtu.edu.cn; yunlong.lu@ieee.org; ruisi.he@bjtu.edu.cn; 23115021@bjtu.edu.cn; chengli@bjtu.edu.cn; boai@bjtu.edu.cn).

Y. Ma is also with the National Mobile Communications Research Laboratory, Southeast University, Nanjing 210096, China.

D. Fei is also with Nanjing Rongcai Transportation Technology Research Institute Co., Ltd, Nanjing, China.

Z. Song is with the Hangzhou Dianzi University, Hangzhou 310018, China. (email: songzh@hdu.edu.cn. )}}

\markboth{}%
{How to Use the IEEEtran \LaTeX \ Templates}

\maketitle

\begin{abstract} 
As next-generation wireless communication systems need to be able to operate in high-frequency bands and high-mobility scenarios, delay-Doppler (DD) domain multicarrier (DDMC) modulation schemes, such as orthogonal time frequency space (OTFS), demonstrate superior reliability over orthogonal frequency division multiplexing (OFDM). Accurate DD domain channel modeling is essential for DDMC system design. However, since traditional channel modeling approaches are mainly confined to time, frequency, and space domains, the principles of DD domain channel modeling remain poorly studied. To address this issue, we propose a systematic DD domain channel measurement and modeling methodology in high-speed railway (HSR) scenarios. First, we design a DD domain channel measurement method based on the long-term evolution for railway (LTE-R) system. Second, for DD domain channel modeling, we investigate quasi-stationary interval, statistical power modeling of multipath components, and particularly, the quasi-invariant intervals of DD domain channel fading coefficients. Third, via LTE-R measurements at 371 km/h, taking the quasi-stationary interval as the decision criterion, we establish DD domain channel models under different channel time-varying conditions in HSR scenarios. Fourth, the accuracy of proposed DD domain channel models is validated via bit error rate comparison of OTFS transmission. In addition, simulation verifies that in HSR scenario, the quasi-invariant interval of DD domain channel fading coefficient is on millisecond (ms) order of magnitude, which is much smaller than the quasi-stationary interval length on $100$ ms order of magnitude. This study could provide theoretical guidance for DD domain modeling in high-mobility environments, supporting future DDMC and integrated sensing and communication designs for 6G and beyond.
\end{abstract}


\section{Introduction}
\label{section:1}
\IEEEPARstart{I}{n} the sixth generation communication system (6G), the scale of high-mobility communication scenarios is expected to dramatically increase \cite{0,M4,Add3}, such as high-speed railway (HSR) communication \cite{LTE-R1}, vehicle-to-everything (V2X) \cite{M3,M11,LiChaoyi}, low-altitude wireless network \cite{6G1} and low earth orbit (LEO) satellite communications \cite{ll2}, which puts forward higher requirements for system transmission reliability. In the 4G and 5G communication systems, the orthogonal frequency division multiplexing (OFDM) waveform has been widely used \cite{M7,OFDM1}. However, its performance will degrade significantly in the high-speed scenarios required by 6G with maximum velocity up to 1000 km/h \cite{6G2,6G3,6G5}. This is because high-speed mobility causes significant Doppler spread, which will destroy the orthogonality of subcarriers and generates inter-carrier interference (ICI) \cite{ICI1}. 

To address this challenge, the delay-Doppler (DD) domain multicarrier modulation (DDMC) schemes were proposed, such as the orthogonal time-frequency space (OTFS) modulation scheme \cite{Hadani2017}. As innovative technologies based on DD domain multiplexing, DDMC schemes can obtain full time-frequency domain diversity gain based on the DD domain channel characteristics, thereby achieving better transmission reliability than OFDM in high-speed mobile scenarios \cite{OTFS1,OTFS2,DDMeasure,Add2}. Accurately understanding the characteristics of the DD domain channel is thus essential for DDMC system design. Different from the doubly-selective time-frequency (TF) domain channel, the DD domain channel exhibits quasi-stationarity, sparsity, separability, and compactness \cite{M5,M6,ChannelEstimation3}: the multipath fading coefficients in the channel stationary interval are almost unchanged; the number of real multipaths is much smaller than the number of system-resolvable multipaths; multipaths generated by different scatterers can be distinguished in the DD domain; and the product of the maximum delay and maximum Doppler of multipaths is much smaller than 1. To understand the characteristics of DD domain channels, DD domain channel measurement and modeling has attracted research attentions.

For DD domain channel measurement, \cite{M11,M12,M13} have conducted studies in V2X and train communication scenarios. Reference \cite{M11} detailed a self-developed millimeter-wave channel detector for V2X, which used all-pilot OFDM and two-dimensional Fourier transform for DD domain channel measurement, confirming the channel’s high sparsity. Reference \cite{M12} proposed an OFDM-based DD domain channel parameter extraction method and revealed the mapping between multipath distribution and scattering environment. Reference \cite{M13} obtained the measured data of the train DD domain channel under different scenarios, including hilly regions and stations. It can be observed that current mainstream methods typically rely on active channel sounding using dedicated all-pilot OFDM waveforms to obtain the full channel response. However, such active methods incur high deployment costs and are difficult to implement on operational HSR base stations. Furthermore, the impact of ICI on DD domain channel characteristics within these full-band OFDM sounding systems remains insufficiently analyzed.

% It can be observed that the current mainstream method first measures the TF domain channel with all-pilot OFDM, then converts it to DD domain for parameter estimation. However, this system cannot avoid ICI of the OFDM system which limits measurement accuracy.

For DD domain channel modeling, reference \cite{M14} built a 60 GHz V2X vehicle channel model for urban intersections, evaluating DD domain amplitude distribution via local scattering functions based on TF domain measurements and orthogonal basis functions. Reference \cite{M15} performed time-delay domain modeling for the Naples-Rome railway (40 km/h train speed), verified DD domain multipath scattering over long observation periods to analyze the stability of signal propagation. Reference \cite{M16} addressed underwater acoustic DD domain measurement, proposed a generative adversarial network-based modeling method, and verified feasibility by comparing coherence time and coherence bandwidth to the measurement results. Reference \cite{ChannelEstimation1} analyzed DD domain channel characteristics and evaluated OTFS performance in HSR scenarios using measured data. Based on the above, the current DD domain modeling research focuses on the mapping between DD domain amplitude distribution, environmental evolution, and multipath birth-death over long periods, but pays little attention to statistical modeling of DD domain multipath fading coefficients, delay, and Doppler in quasi-stationary intervals.

For HSR channel modeling, current research is mainly carried out in the time domain, i.e., modeling and analyzing the time-varying channel impulse response (CIR) \cite{ActiveMeasure3-Modeling1,PassiveMeasure1-Modeling3,DingJianwen,YangJingya,ActiveMeasure6-Modeling2}. Reference \cite{RTModeling1} developed ray tracing-based simulation models to investigate radio wave propagation behaviors in HSR scenarios. Reference \cite{GaotieModeling1} proposed a nonstationary geometry-based stochastic model for wideband MIMO HST channels in rural macrocell scenarios, and a simulation model via the modified equal-area method. In reference \cite{ActiveMeasure5-GaojiaqiaoModeling1}, researchers simulated and analyzed the channel in viaduct scenarios based on measured data. References \cite{mmwaveActiveMeasure1-CDLModeling1,YangMi} performed multipath clustering and tracking on the multipath components (MPCs) of the channel measurement results, and conducted statistical modeling on inter-cluster and intra-cluster parameters respectively. Despite the progress of these studies in characterizing HSR channel propagation, correlation, and fading coefficients, existing HSR channel research has paid limited attention to the channel characteristics in the DD domain, resulting in a notable research gap in this area.

Based on the above analysis, we note that in the HSR scenario, the statistical distribution of DD domain channel parameters within the quasi-stationary interval has received little attention, and there is a lack of systematic DD domain channel measurement and modeling theory, which restricts the frontier research and implementation of DDMC. To fill this gap, in this paper, we propose a systematic DD domain channel measurement and modeling methodology for the HSR scenario. The main innovations of this work include:

\begin{itemize}
	\item We propose a DD domain channel measurement method based on the LTE-R framework. Unlike active sounding methods that require dedicated waveforms, we utilize discrete TF domain pilots (cell specific reference signal (CRS)) within the existing LTE-R standard. Therein, to circumvent the errors caused by traditional TF domain interpolation, a coarse estimation of DD domain channel parameters is first performed, followed by the proposal of a precise estimation method for delay, Doppler shift, and fading coefficient, thereby enabling commercial OFDM systems to conduct DD domain channel measurements.
	\item We propose the quasi-stationary DD domain channel modeling method. In detail, the DD domain channel quasi-stationarity evaluation method is first designed, the statistical modeling of DD domain channel parameters is then investigated within quasi-stationary intervals, and finally the quasi-invariant interval of channel fading coefficient is modeled to guide the design of DDMC schemes.
	\item Based on the proposed DD domain channel measurement and modeling method, taking the quasi-stationary interval as the decision criterion, we establish DD domain channel models under different channel time-varying conditions in HSR viaduct scenarios. The models provide the quasi-stationary interval of the DD domain channel, the delay of multipath, Doppler shift, amplitude power distribution, and the minimum quasi-invariant interval of each path.
	\item We validate the fidelity of the proposed models through a multi-dimensional verification framework. Specifically, we verify the consistency of physical dispersion parameters (root mean square (RMS) delay and Doppler spreads) and demonstrate the model's robustness under high-order modulations, ensuring both physical and statistical accuracy, which indicates that the established DD domain channel models accurately capture the characteristics of realistic channels. Additionally, we verify that the DD domain channel can be regarded as invariant when the smallest quasi-invariant interval of multipath is satisfied, which is on the ms order in the HSR viaduct scenario.
\end{itemize}

The paper is organized as follows: Section \ref{section:2} introduces the LTE-R channel measurement method, and channel parameter acquisition method based on discrete TF domain pilots. Section \ref{section:3} presents the quasi-stationary DD domain channel modeling method. Section \ref{section:4} describes the DD domain channel models under different stationary intervals in HSR viaduct scenarios. Finally, Section \ref{section:5} concludes this work and outlines further researches.

\section{LTE-R Based DD Domain Channel Measurement}
\label{section:2}
LTE-R is a specialized variant of LTE technology designed for railway communications \cite{LTE-R1}. This section first introduces the measurement settings of LTE-R, then presents coarse DD domain channel estimation method based on discrete TF domain pilots, and finally proposes a precise DD domain channel parameters estimation method.

\subsection{Measurement Settings}
\label{section:2-1}
\begin{figure}[t]
	\centering
	\includegraphics[width=\linewidth]{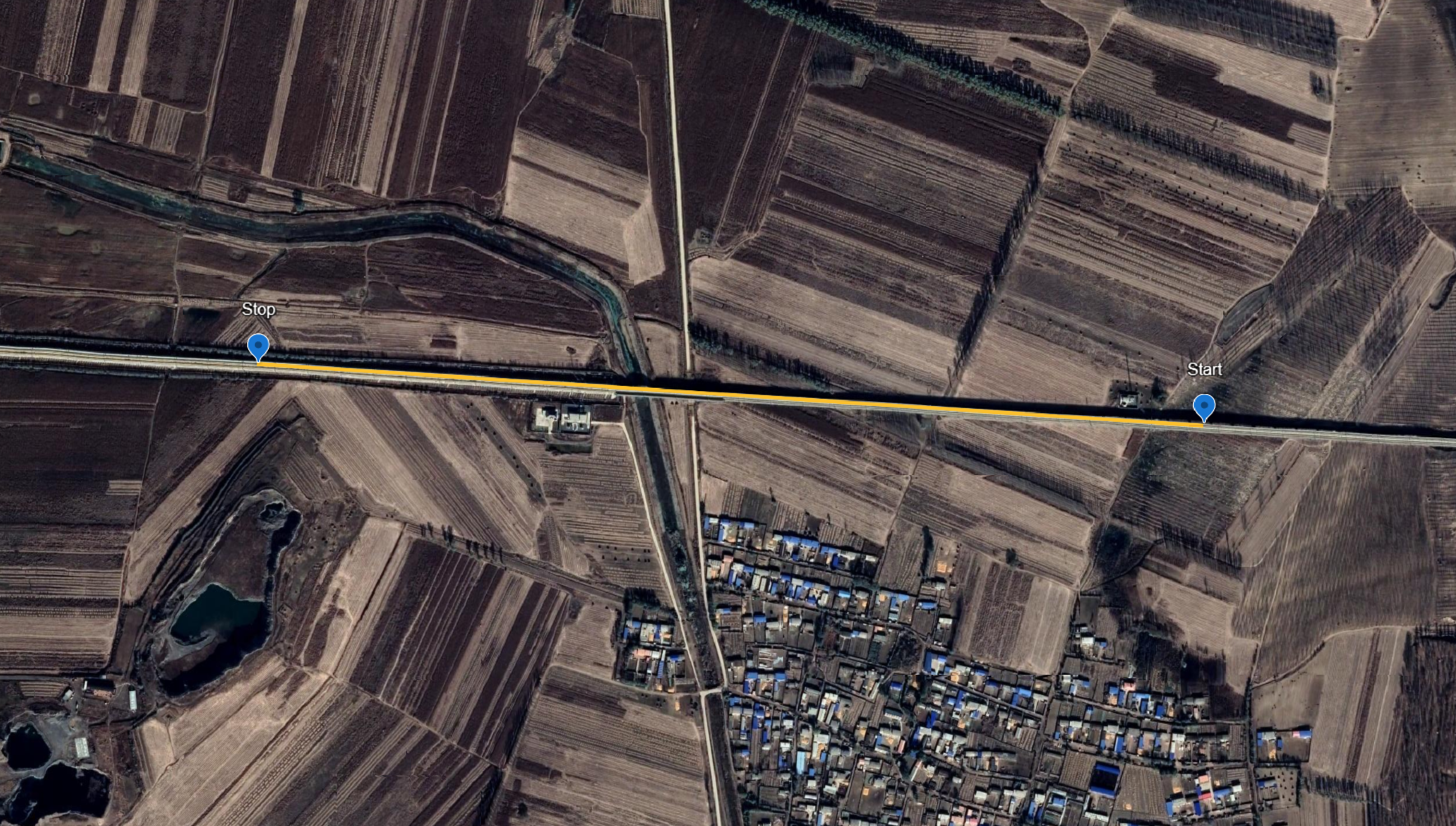}
	\caption{Measurement scenario diagram.}
	\label{fig1}
\end{figure}

We conducted HSR channel measurement in the viaduct at a rural area along the Beijing-Shenyang HSR in China, and the measurement scenario is shown in Fig. \ref{fig1}. The distance from the starting point to the end point is approximately 1450 m, and the area around the line is mostly farmland and low-rise buildings. The starting and ending points of the measurement are marked in blue. 

To receive LTE-R signals, we modify an universal software radio peripheral (USRP) according to the LTE-R standard to act as the channel sounder. The USRP is equipped with 2 receiving antennas and GPS clocks. During measurement, the channel sounder keeps receiving the physical downlink control channel (PDCCH) signal sent by the LTE-R base stations along the rail. The parameter settings of the measurement system are shown in Table \ref{tab1:parameters}, same as those in \cite{ChannelEstimation1}.

\begin{table}[h]
	\centering
	\caption{System Parameter Configuration}
	\label{tab1:parameters}
	\renewcommand{\arraystretch}{1.5}
	\begin{tabular}{p{3.5cm}p{3cm}}
		\hline
		\textbf{Parameters} & \textbf{Setting} \\
		\hline
		Frequency & 465 MHz \\
		
		Bandwidth & 5 MHz \\
		
		Transmission Mode & LTE FDD \\
		
		Subcarrier Spacing & 15 kHz \\
		
		Transmit Power & 43 dBm \\
		
		Tx Antenna Gain & 12 dBi \\
		
		Rx Antenna Gain & 4 dBi \\
		
		Sample Rate & 7.68 MHz \\
		
		HSR Speed & 371 km/h \\
		
		Antenna Height & 4.05 m \\
		
		BS Height & about 40 m \\
		\hline
	\end{tabular}
\end{table}
\subsection{TF Domain Channel Fading Acquisition}
\label{section:2-2}
Through LTE-R channel measurement, the I/Q components of received signals and the corresponding geographic information are collected, including timestamps, position information,  and the train speed. Afterwards, the TF domain channel fading corresponding to discrete pilots need to be extracted. In detail, the raw data will be preprocessed in the following steps:

\subsubsection{Cell ID Detection and Frame Offset Correction} Let \( r(t) \) represent the received continuous time-domain signal, and \( r[n] \) be its sampled version. The cell search process can be expressed as:
\begin{equation}
	(N_{\text{ID}}^{\text{cell}}, \Delta_{\text{frame}}) = \underset{N_{\text{ID}}^{\text{cell}}, \Delta_{\text{frame}}}{\text{argmax}} \left| \sum_{k} r[k+\Delta_{\text{frame}}] \cdot s_{N_{\text{ID}}}^{*}[k] \right|,
\end{equation}
where \( N_{\text{ID}}^{\text{cell}} \) is the detected cell ID, \( \Delta_{\text{frame}} \) is the frame offset, \( s_{N_{\text{ID}}}[k] \) is the synchronization signal sequence corresponding to cell ID \( N_{\text{ID}} \), \( ^* \) denotes conjugation. Afterwards, the frame start position will be corrected based on the estimated frame offset \( \Delta_{\text{frame}} \).

\subsubsection{Resource Grid Acquisition} Then, the synchronization signals will be removed according to the LTE-R standard and cell ID. After that, the time domain signal will be transformed to the frequency domain to obtain the received symbols on each subcarrier, forming a resource grid. This process
includes removing the cyclic prefix (CP) and performing fast Fourier transform (FFT). For the \( n \)-th OFDM symbol, its time domain representation (after removing CP) is denoted as \( r_n[q] \). The frequency domain resource element (RE) \( Y_n[m] \) (the \( m \)-th subcarrier of the \( n \)-th symbol) can be obtained via FFT:
\begin{equation}
	Y_n[m] = \sum_{q=0}^{N_{\text{FFT}}-1} r_n[q] e^{-j2\pi qm/N_{\text{FFT}}},
\end{equation}
where \( N_{\text{FFT}} \) is the number of FFT points, $n\in\{0,1,...,N-1\}, m \in \{0,1,...,M-1\}$, and $N$ and $M$ are the number of OFDM symbols and subcarriers, respectively.

Subsequently, the TF domain channel fading corresponding to the CRS will be estimated. Assume the resource grid has \( M \) subcarriers in the frequency domain and \( N \) OFDM symbols in the time domain. Let \( Y[m,n] \) be the received signal on the \( m \)-th subcarrier and \( n \)-th OFDM symbol, and \( X_p[m,n] \) be the corresponding transmitted pilot symbol. According to reference \cite{article11,ma2025delay}, the relationship between $Y[m,n]$ and $X_p[m,n]$ is given by:
\begin{equation}
	\label{Y(m,n)}
	 Y[m,n] = H[m,n]X_p[m,n] + I[m,n] + w[m,n],
\end{equation} 
where $H(m,n)$ refers to the channel fading of $X_p[m,n]$ in $Y[m,n]$, $I[m,n]$ refers to ICI experienced by $Y[m,n]$, and $w[m,n]$ denotes noise. In the frequency domain, the ICI term $I[m,n]$ arises from the superposition of leakage components from multiple subcarriers. According to the central limit theorem (CLT), this aggregated interference can be approximated as a complex Gaussian random variable \cite{CLT}. Consequently, the ICI behaves similarly to additive noise. By treating the ICI as part of the total noise and utilizing the least square (LS) criterion, the initial TF domain channel fading estimates at pilot positions can be obtained as:
\begin{equation}
	\small
	\hat{H}_{LS}[m,n] = \frac{Y[m,n]}{X_p[m,n]} = H[m,n] + \frac{I[m,n] + w[m,n]}{X_p[m,n]}.
\end{equation}\par
After obtaining the initial estimates, we must isolate the pilot information from the rest of the frame. In the LTE downlink frame structure, the resource grid is complex. Besides the CRS used for channel estimation, the grid is populated by the physical downlink control channel (PDCCH), the physical downlink shared channel (PDSCH), and synchronization signals (PSS/SSS). For the received signal $Y[m,n]$ at non-pilot locations (i.e., $[m,n] \notin \mathcal{P}$), the value represents data or control information superimposed with noise, rather than zero. Using these values directly would introduce significant interference to the DD channel estimation. To address this, we perform an algorithmic masking operation (zero-padding). We retain the LS channel estimates at the CRS positions and artificially set the values at all other resource elements to zero. This process filters out the interference from PDSCH and PDCCH, resulting in the TF channel matrix $\hat{H}_{\text{LS}}[m,n]$:
\begin{equation}
	\label{TF domain channel fading}
	\hat{H}_{\text{LS}}[m,n] = 
	\begin{cases} 
		\frac{Y[m,n]}{X_p[m,n]} & \text{if } [m,n] \in \mathcal{P} \\
		0 & \text{if } [m,n] \notin \mathcal{P},
	\end{cases}
\end{equation}
where \( \mathcal{P} \) represents the set of all pilot symbol positions in the time-frequency resource grid. Then, the obtained $\hat{H}_{\text{LS}}[m,n]$ will be used for subsequent DD domain channel estimation and modeling.

\subsection{Coarse Estimation of DD Domain Channel}
\label{section:2-3}

First, the DD domain channel can be modeled as \cite{article11,ma2025delay}:
\begin{equation}
	\label{h(tau,nu)}
	h(\tau,\nu) = \sum\limits_{i=1}^{P} h_i  \delta(\tau-\tau_i)\delta(\nu-\nu_i),
\end{equation}
where $\tau\in [0,\tau_\text{max}]$, $\nu\in [-\nu_\text{max},\nu_\text{max}]$, and $h_i$ denote delay, Doppler, and fading coefficient of $i$-th multipath, respectively, $P$ represents the number of multipath, and $\tau_\text{max}$ and $\nu_\text{max}$ are maximum delay and Doppler, respectively. The normalized delay $l_i$ and Doppler $k_i$ can be expressed as \cite{article11,ma2025delay}:
\begin{equation}
	\label{li,ki}
	l_i = M\Delta f \tau_i, k_i = NT \nu_i.
\end{equation}

A key challenge in extracting channel parameters from LTE-R signals is ICI caused by high speed train movement. However, in the context of DD domain extraction, the impact of ICI can be effectively managed. Since the transformation from the TF domain to the DD domain involves unitary operations, the statistical properties of Gaussian noise are preserved. Therefore, the ICI term, which is modeled as Gaussian noise in the TF domain, remains Gaussian-distributed in the DD domain: $I_{DD}[k,l] \sim \mathcal{CN}(0, \sigma_{ICI}^2).$ Unlike the channel response energy, which is concentrated in sparse multipath peaks in the DD domain, the energy of ICI is spread across the entire Delay-Doppler plane. This results in an increase in the effective noise floor $\sigma_{eff}^2 = \sigma_{w}^2 + \sigma_{ICI}^2$. As will be introduced in Section \ref{section:2-4}, we adopt a threshold-based mechanism where valid multipaths must exceed the effective noise floor by 6 dB, ensuring that ICI functions as background noise and does not obscure the extraction of dominant channel features. Based on (\ref{Y(m,n)}), (\ref{TF domain channel fading}), and the fact that $H[m,n]$ in (\ref{Y(m,n)}) can be written as $H[m,n] = \sum^P_{i=1}h_ie^{j2\pi\frac{nk_i}{N}}e^{-j2\pi\frac{ml_i}{M}}$ according to (\ref{li,ki}), the coarse estimation of the DD domain channel can be obtained through applying FFT along the time domain and the IFFT along the frequency domain on (\ref{TF domain channel fading}), leading to \cite{article11,ma2025delay}:
\begin{equation}
	\begin{array}{l}
		{\hat h}_{\rm DD}[k,l] = FF{T_N}\{ IFF{T_M}\{\hat{H}_{\text{LS}}[m,n]\} \} \\
		= \sum\limits_{i = 1}^P {{h_i}} \sum\limits_{m = 0}^{M - 1} {\frac{1}{{\sqrt M }}{e^{ - j2\pi m\frac{{({l_i} - l)}}{M}}}} \sum\limits_{n = 0}^{N - 1} {\frac{1}{{\sqrt N }}{e^{ j2\pi n\frac{{({k_i} - k)}}{N}}}},
	\end{array}
\end{equation}
where $k\in\{0,1,...,N-1\}, l \in \{0,1,...,M-1\}$. Besides, we denote that pilot spacing of CRS is $d_t$ in time domain and $d_f$ in frequency domain, with ${\rm mod}(M,d_f) = 0$ and ${\rm mod}(N,d_t) = 0$. We note that $\hat h_{\rm DD}[k,l]$ is two-dimensional periodic with periods $\frac{M}{{{d_f}}}$ and $\frac{N}{{{d_t}}}$ over delay and Doppler domains, and thus $\hat h_{\rm DD}[k,l]$ is recorded as ${\hat h}_{\rm DD}^{\rm Period}[k,l]$ \cite{article11,ma2025delay}:
\begin{equation}
	\label{hDDPeriod}
	\begin{array}{l}
		{\hat h}_{\rm DD}^{\rm Period}[k,l] = \\
		\sum\limits_{i = 1}^P {{h_i}} \underbrace {\sum\limits_{m' = 0}^{\frac{M}{{{d_f}}} - 1} {\frac{{{d_f}}}{{\sqrt M }}{e^{ - j2\pi m'{d_f}\frac{{({l_i} - l)}}{M}}}} }_{{\cal C}^{\rm Period}_{\rm Delay}({l_i},l)}\underbrace {\sum\limits_{n' = 0}^{\frac{N}{{{d_t}}} - 1} {\frac{{{d_t}}}{{\sqrt N }}{e^{j2\pi n'{d_t}\frac{{({k_i} - k)}}{N}}}} }_{{\cal C}^{\rm Period}_{\rm Doppler}({k_i},k)},
	\end{array}
\end{equation}
where ${\cal C}^{\rm Period}_{\rm Delay}({l_i},l)$ and ${\cal C}^{\rm Period}_{\rm Doppler}({k_i},k)$ represent the delay and Doppler sampling components. However, $\hat h_{\rm DD}^{\rm Period}[k,l]$ in (\ref{hDDPeriod}) is the sampled version of the physical DD domain channel in (\ref{h(tau,nu)}). For DD domain channel modeling, precise parameters of the physical DD domain channel are desired.

\subsection{Precise DD Domain Channel Estimation}
\label{section:2-4}

The coarse estimation relies on the direct symplectic finite Fourier transform to convert the TF domain estimate into the DD domain. This direct transformation is fundamentally limited by the grid resolution determined by the pilot mode. In high mobility scenarios such as HSR, physical multipath components are continuous and often located off-grid, leading to severe energy leakage and aliasing on the double-difference grid, problems that cannot be solved by a simple inverse transform. To address this, we propose an precise DD domain channel estimation in this section, including multipath delays, Doppler shifts, and path powers based on $l_i,k_i,h_i$ in (\ref{hDDPeriod}). This precise DD domain channel estimation method effectively decouples the estimation accuracy from the pilot grid resolution, achieving super-resolution performance that direct transformation methods cannot achieve.

To the beginning, the valid MPCs need to be extracted from ${\cal C}^{\rm Period}_{\rm Delay}({l_i},l)$ in (\ref{hDDPeriod}). In detail, we  use an energy threshold to identify valid MPC, where the threshold $Th$ is set to be 6 dB higher than the noise floor according to the classical channel modeling literature \cite{article10}. Then, a MPC is regarded as valid if its power satisfies $|{\hat h}_{\rm DD}^{\rm Period}[k,l]|^2  > Th$ and $|{\hat h}_{\rm DD}^{\rm Period}[k,l]|$ is a peak value. The number of valid MPC is denoted as $\hat P$. Besides, the indices of the valid MPC's delay and Doppler are collected as $\{ k_i, l_i\}$, where $i$ is number of multipath. Therein, we assume the delays or Doppler shifts of valid MPCs are different. Afterwards, we adopt Algorithm 1 to jointly estimate the off-grid delay \( \hat{l}_i \), Doppler shift \( \hat{k}_i \), and corresponding fading coefficient \( \hat{h}_i \) of the physical DD domain channel according to \cite{article11,ma2025delay,article12}.

\begin{algorithm}[h]
	\caption{Physical DD domain channel estimation}
	\label{algo:FDD}
	\begin{algorithmic}[1]
		\Require Estimated ${\hat{h}_{\rm{DD}}^{\rm{Period}}[k,l]}$, Number of Multipath $\hat{P}$
		\Ensure Channel parameters $\hat{l}_i$, $\hat{k}_i$ and $\hat{h}_i$ for $i \in \{1, \ldots, \hat{P}\}$ and estimated DD domain channel \(\hat h_{{\rm{DD}}}^{{\rm{Est}}}[k,l]\)
		%\For{$a = d_t$ to $n \times d_t$}
		\State $\hat h_{{\rm{DD}}}^{{\rm{Est}}}[k,l] = 0$, $k\in\{0,1,...,N-1\}, l \in \{0,1,...,M-1\}$
		\For{$i = 1$ to $\hat{P}$}
		\State $[k_0, l_0] = \{ k_i, l_i \}$
		\State $l_1 = \underset{l \in \{l_0 - 1, l_0 + 1\}}{\mathrm{argmax}} |{\hat{h}_{\rm{DD}}^{\rm{Period}}[k_0,l]}|$
		\State $k_1 = \underset{k \in \{k_0 - 1, k_0 + 1\}}{\mathrm{argmax}} |{\hat{h}_{\rm{DD}}^{\rm{Period}}[k,l_0]}|$
		\State $\hat l_i = l_0 + \frac{|\hat{h}_{\rm{DD}}^{\rm{Period}}[k_0, l_1]| |(l_1 - l_0) |}{|\hat{h}_{\rm{DD}}^{\rm{Period}}[k_0, l_0]| + |\hat{h}_{\rm{DD}}^{\rm{Period}}[k_0, l_1]|}$
		\State $\hat{k}_i = k_0 + \frac{|\hat{h}_{\rm{DD}}^{\rm{Period}}[k_1, l_0]| |(k_1 - k_0)|}{|\hat{h}_{\rm{DD}}^{\rm{Period}}[k_0, l_0]| + |\hat{h}_{\rm{DD}}^{\rm{Period}}[k_1, l_0]|}$
		\State $\hat{h}_i = \frac{\hat{h}_{\rm{DD}}^{\rm{Period}}[k_0, l_0]}{\mathcal{C}^{\rm{Period}}(\hat{l}_i, l_0)\mathcal{C}^{\rm{Period}}(\hat{k}_i, k_0)}$
		\State \small $\hat h_{{\rm{DD}}}^{{\rm{Rebuild}}}[k',l'] = {\hat h_i} {{\cal C}^{{\rm{Period}}}}({\hat l_i},{l_0}){{\cal C}^{{\rm{Period}}}}({\hat k_i},{k_0})$
		\State $\hat h_{{\rm{DD}}}^{{\rm{Period}}}[k,l] = \hat h_{{\rm{DD}}}^{{\rm{Period}}}[k,l] - \hat h_{{\rm{DD}}}^{{\rm{Rebuild}}}[k',l']$
		\State $\hat h_{{\rm{DD}}}^{{\rm{Est}}}[k,l] = \hat h_{{\rm{DD}}}^{{\rm{Est}}}[k,l] + \hat h_{{\rm{DD}}}^{{\rm{Rebuild}}}[k',l']$
		\EndFor
		%\State $j = a/d_t$
		\State Save $\{\hat{h}_i, \hat{k}_i, \hat{l}_i\}$ for $\hat{P}$ multipath and $\hat h_{{\rm{DD}}}^{{\rm{Est}}}[k,l]$.
		%\EndFor
	\end{algorithmic}
\end{algorithm}

As shown in Algorithm 1, in addition to ${\hat{h}_{\rm{DD}}^{\rm{Period}}[k,l]}$ in (\ref{hDDPeriod}), we will input the estimation of the number of multipath $\hat{P}$. From Steps 3 to 7, the indices of physical Dopplers and delays, i.e., $\hat{k}_i$ and $\hat{l}_i$, are estimated. First, the sub-max value of ${\hat{h}_{\rm{DD}}^{\rm{Period}}[k_0,l]}$ is detected for $l \in\{l_0 - 1, l_0 + 1\}$, and the corresponding index of delay is denoted as $l_1$ in Step 4. Then, based on the relationship of the max and sub-max values of the function $sin(x)/sin(x/N)$ when $N$ is large \cite{article11,ma2025delay,article12}, we can estimate the physical delay $\hat{l}_i$ in Step 6. Similarly, the physical Doppler $\hat{k}_i$ can be estimated in Step 5 and Step 7. Next, in Step 8, the channel fading coefficient $\hat{h}_i$ is estimated based on $\hat{k}_i, \hat{l}_i$, and (\ref{hDDPeriod}). In Step 9, we obtain $\hat h_{{\rm{DD}}}^{{\rm{Rebuild}}}[k',l']$ of the $i$-th path based on $\hat{k}_i, \hat{l}_i, \hat{h}_i$. In Step 10, the newly estimated $\hat h_{{\rm{DD}}}^{{\rm{Rebuild}}}[k',l']$ is subtracted from the original $\hat h_{{\rm{DD}}}^{{\rm{Periodic}}}[k,l]$ to eliminate multipath interference. Finally, $\{\hat{h}_i, \hat{k}_i, \hat{l}_i\}$ for $\hat{P}$ multipath are obtained.

In summary, based on the above methods, the complete process of DD domain channel measurement can be summarized as following steps: First, the TF domain channel fading $\hat{H}_{\text{LS}}[m,n]$ is acquired from HSR measured data. Then, the DD domain channel coarse estimation method is used to obtain ${\hat h}_{\rm DD}[k,l]$ and ${\hat h}_{\rm DD}^{\rm Period}[k,l]$. Finally, the precise estimation method is employed to obtain the DD domain channel delay $\hat{l}_i$, Dopper $\hat{k}_i$ and fading coefficient $\hat{h}_i$.

\section{LTE-R Based DD Domain Channel Modeling}
\label{section:3}

This section focuses on the DD domain channel modeling methods, with emphasis on the quasi-stationary interval evaluation, the multipath amplitude distribution modeling, and the quasi-invariance evaluation methods.

\subsection{Quasi-Stationary Interval Evaluation}
\label{section:3-1}

According to \cite{WSSUS}, \cite{Molisch2005}, for a wide-sense stationary uncorrelated scattering (WSSUS) channel, the channel correlation in the time-delay domain can be given as:
\begin{equation}
	\mathbb{E}\{h(t,\tau)h^*(t',\tau')\} = r_h(t - t';\tau)\delta(\tau - \tau'),
\end{equation}
with $r_h(\Delta t;\tau)$ being the correlation function of time difference $\Delta t = t - t'$ and delay $\tau$. In DD domain, we define the \textit{scattering function} $S(\tau, \nu)$ as the Fourier transform of $r_h(\Delta t;\tau)$ with respect to $\Delta t$:
\begin{equation}
	S(\tau, \nu) = \int_{-\infty}^{\infty} r_h(\Delta t;\tau) e^{-j2\pi\nu \Delta t} d(\Delta t),
\end{equation}
where $S(\tau, \nu)$ represents the power density at delay $\tau$ and Doppler shift $\nu$, describing the scattered signal power at $(\tau, \nu)$. Therefore, the DD domain channel correlation for the WSSUS channel is given by:
\begin{equation}
	\label{DD-WSSUS}
	\mathbb{E}\{h(\tau,\nu)h^*(\tau',\nu')\} = S(\tau, \nu) \delta(\tau - \tau') \delta(\nu - \nu').
\end{equation}

This shows that in the stationary intervals, the delay and Doppler shift of multipaths in the DD domain do not change significantly, and no multipath birth or death occurs. Thus, we can consider that the statistical performance of multipaths remains identical within a stationary interval. Based on (\ref{DD-WSSUS}), we use a spectral distance metric to quantify the channel stationarity, called collinearity \cite{article9}. The collinearity is a bounded metric $\gamma \in [0,1]$ that compares the DD domain at different time instances. A collinearity close to 1 is an outcome for very similar power spectra, whereas a collinearity close to 0 results from comparing two very dis-similar spectral densities. The CDD in time is defined as:
\begin{equation}
	\label{CDD}
	\small
	CDD(t_i,t_j) = \frac{{\sum\limits_{p = 0}^{M - 1} {\sum\limits_{q = 0}^{N - 1} {P({t_{i,}}{\tau _p},{\nu _q})P({t_{j,}}{\tau _p},{\nu _q})} } }}{{\sqrt {\sum\limits_{p = 0}^{M - 1} {\sum\limits_{q = 0}^{N - 1} {P{{({t_{i,}}{\tau _p},{\nu _q})}^2}} } } \sqrt {\sum\limits_{p = 0}^{M - 1} {\sum\limits_{q = 0}^{N - 1} {P{{({t_{j,}}{\tau _p},{\nu _q})}^2}} } } }},
\end{equation}
where \({P(t_{i}, \tau_p, \nu_q)}\) represents the power of $h(\tau,\nu)$ at time \(t_{i}\), delay \(\tau_p\), and Doppler shift \(\nu_q\) with $\tau_p = \frac{p}{M\Delta f}$, $\nu_q = \frac{q}{NT}$. By constructing a CDD matrix, quasi-stationary intervals $T_{QS}$ in the DD domain can be identified through thresholding. 

To quantify the channel stationarity, we utilize the CDD metric defined in (\ref{CDD}). The evaluation process primarily involves the selection of a similarity threshold and the statistical analysis of stationary durations. First, the choice of the similarity threshold $\alpha$ is critical for defining the stationarity region. Following established practices in recent channel measurement literature \cite{TPCC,Threhold2}, we select the range $\alpha \in [0.7, 0.9]$. 
	
To ensure the validity of the measurement results and the high fidelity of the derived channel models, we prioritize the higher threshold of $\alpha = 0.9$ for the final parameter extraction and modeling presented in this work. As suggested in \cite{TPCC}, a conservative threshold is essential for high-fidelity modeling to rigorously satisfy the wide-sense stationary (WSS) assumption and minimize the inclusion of transient non-stationary states. While a lower threshold (e.g., $\alpha = 0.7$) is commonly utilized as the minimum validity bound for WSS in highly dynamic vehicular environments \cite{article9}, relying on it may introduce statistical estimation errors due to the inclusion of weakly stationary segments. Therefore, we adopt 0.9 as the primary decision criterion to guarantee model accuracy, while using the range down to 0.7 to verify robustness.

Based on the calculated CDD values and the chosen threshold, we identify the stationary regions along the diagonal of the correlation matrix. Let $\mathcal{T}_k = \{t_{k, \text{start}}, \dots, t_{k, \text{end}}\}$ denote the $k$-th identified continuous time segment where the CDD between any two time instants within this segment satisfies $CDD(t_i, t_j) \ge \alpha$. The duration of this $k$-th stationary block is calculated as $T_{k} = t_{k, \text{end}} - t_{k, \text{start}}$. Since the HSR channel is dynamic, the stationary duration varies over the measurement track. Therefore, to obtain a representative model parameter, we perform a statistical analysis on all identified valid durations. The final quasi stationary interval $T_{QS}$ reported in this work is defined as the statistical mean of these durations, i.e., $T_{QS} = \mathbb{E}[T_k]$, representing the average timescale over which the channel statistics remain invariant.

\subsection{Amplitude Distribution Fitting}
\label{section:3-2}
Within a stationary interval, the MPCs in the DD domain exhibit stable statistical characteristics as implied by the WSSUS property. To capture these characteristics, we employ a sliding window of size \(M \times N\) along the Doppler dimension with a step size of $d_t$. The number of sliding points is calculated as \(a = \lfloor (T_{\text{max}} - T_{\text{min}})/d_t \rfloor\), where $T_{\text{max}}$ and $T_{\text{min}}$ denotes the maximum and minimum index values of the RE with size $N \times M$ within the stationary interval. Based on Algorithm 1, the estimated physical delay and Doppler shift of multipath within one stationary interval are calculated using weighting coefficients, given by:
\begin{equation}
	\hat l_i^{\rm weighted} = \frac{{\sum\limits_{j = 1}^a {{\hat l_i}^{(j)}\left| {{\hat h_i}^{(j)}} \right|} }}{{\sum\limits_{j = 1}^a {\left| {{\hat h_i}^{(j)}} \right|} }}.
\end{equation}

\begin{equation}
	\hat k_i^{\rm weighted} = \frac{{\sum\limits_{j = 1}^a {{\hat k_i}^{(j)}\left| {{\hat h_i}^{(j)}} \right|} }}{{\sum\limits_{j = 1}^a {\left| {{\hat h_i}^{(j)}} \right|} }}.
\end{equation}
This process refines the estimation of delay and Doppler parameters for each MPC in the DD domain, yielding more precise representations of their central tendencies. 
\begin{table}[h]
	\centering
	\caption{Probability Distribution Model.} 
	\label{tab2:distribution}
	\renewcommand{\arraystretch}{2}
	\begin{tabular}{cc}
		\hline
		\textbf{Distribution} & \textbf{Probability distribution function} \\
		\hline
		Rician & $f(x|s,\sigma)=I_0\left(\frac{x s}{\sigma^2}\right)\frac{x}{\sigma^2}e^{-\left(\frac{x^{2}+s^{2}}{2\sigma^{2}}\right)}$ \\
		
		Rayleigh & $f(x|b)=\frac{x}{b^{2}}e^{\left(\frac{-x^{2}}{2b^{2}}\right)}$ \\
		
		Nakagami & $f(x|\mu,\omega)=2\left(\frac{\mu}{\omega}\right)^{\mu}\frac{1}{\Gamma(\mu)}x^{(2\mu - 1)}e^{-\frac{\mu}{\omega}x^{2}}$ \\
		
		Weibull & $f(x|a,b)=\frac{b}{a}\left(\frac{x}{a}\right)^{b - 1}e^{-\left(\frac{x}{a}\right)^{b}}$ \\
		\hline
	\end{tabular}
\end{table}

While the averaged delay and Doppler parameters effectively characterize the temporal stability of MPCs, the power characteristics require further statistical analysis. Given that MPCs within a stationary interval share identical statistical performance, we focus on fitting probability distributions to the amplitude values of all MPCs within a single data packet captured during this stationary interval. Based on literatures, we select classical probability distributions to fit the amplitude distribution \cite{article12}, i.e., Rician, Rayleigh, Nakagami, and Weibull distributions, as shown in Table \ref{tab2:distribution}.

To quantitatively evaluate the fitting degree of the selected candidate distributions to the observed power data, we will adopt the Kolmogorov-Smirnov (KS) test. The KS test makes a judgment by comparing the maximum absolute difference \( D_n = \sup_x |F_n(x) - F(x)| \) between the empirical cumulative distribution function (ECDF) \( F_n(x) \) of the sample data and the cumulative distribution function (CDF) \( F(x) \) of the theoretical distribution. 

\begin{figure}[!]
	\centering
	\includegraphics[width=\linewidth]{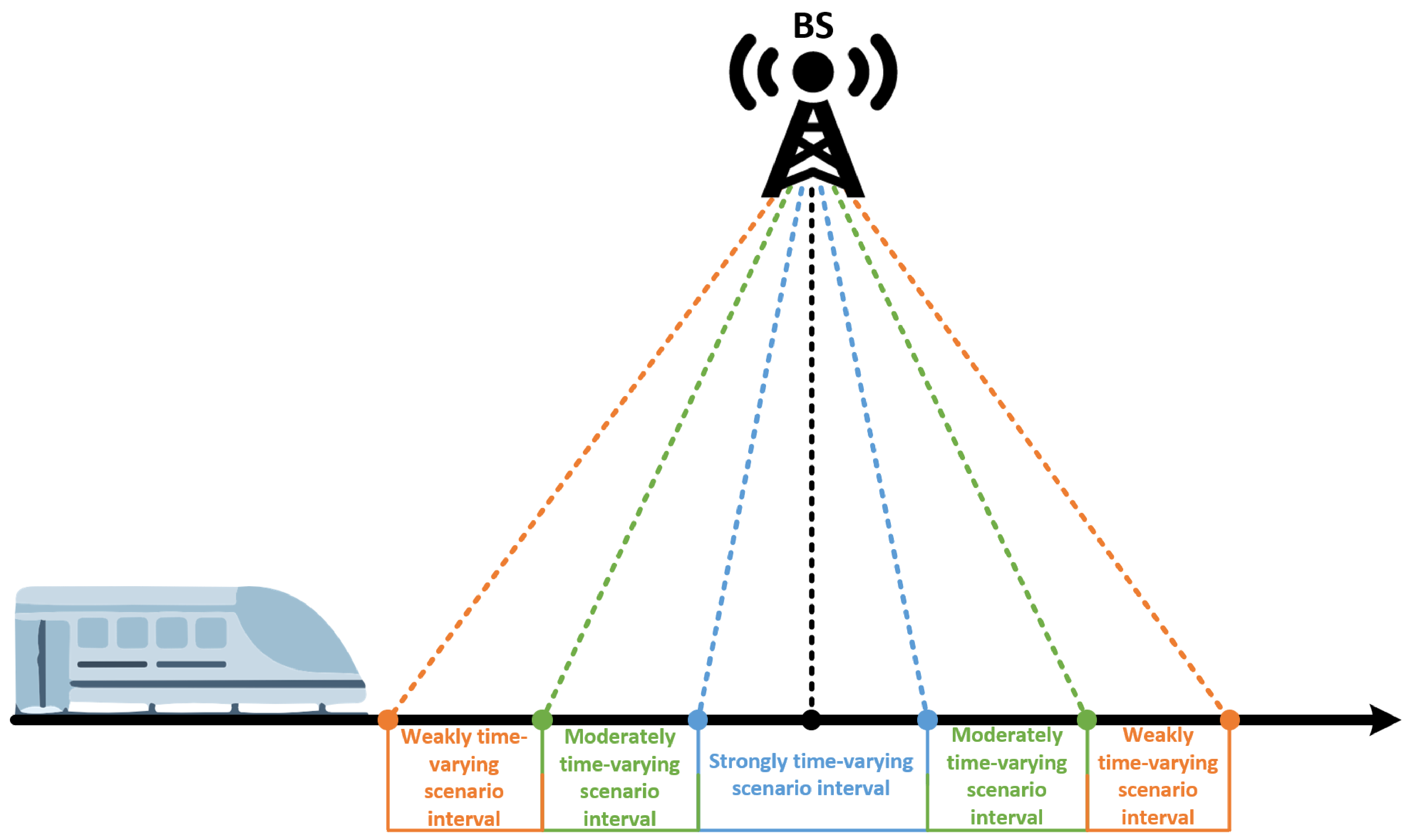}
	\caption{Relative positions over the distance.} 
	\label{fig:Relative Position}
\end{figure}

\subsection{Quasi-Invariant Interval Evaluation}
\label{section:3-3}

The core focus of a stationary interval lies in whether the statistical characteristics of the channel undergo significant changes, rather than the fluctuations of the DD domain channel fading coefficient $h_i$. Even if there are slight random fluctuations in the DD domain channel fading coefficient $h_i$, the interval can still be identified as a stationary interval as long as the statistical characteristics remain stable. However, for DDMC data packets, the variation of $h_i$ is not desired and the excessive fluctuations in $h_i$ may degrade the transmission performance of data.

To address this, we further define a more stringent \textit{quasi-invariant} interval based on the already identified \textit{statistical stationary intervals}. This quasi-invariant interval not only requires the channel to maintain stable statistical characteristics but also imposes constraints on the DD domain channel fading coefficient to ensure no significant changes over time. Specifically, the similarity of instantaneous DD domain fading coefficient will be measured using the DD domain time correlation coefficient (DD-TCC). According to \cite{TPCC}, we extend the time power correlation coefficient (TPCC) from the delay domain to the DD domain, defining the DD-TCC. This coefficient measures each MPC's linear similarity of two DD domain fading coefficient matrices, formulated as:
\begin{equation}
	\label{DD-TCC}
	\small
	\text{DD-TCC}(i, t_b, t_c) = \frac{|h_i(t_b) \cdot h_i^*(t_c)|}{\max\left(|h_i(t_b)|^2, |h_i(t_c)|^2 \right)},
\end{equation}
where $i \in \{1, \ldots, \hat{P}\}$, \( h_i(t_b) \) and \( h_i(t_c) \) denote the DD domain channel fading coefficient of the  $i$-th multipath at the \( b \)-th and \( c \)-th time instants. By constructing a DD-TCC matrix, quasi-invariant intervals $T_{QI}$ in the DD domain can be identified through thresholding. The specific steps are similar to those of quasi-stationary interval modeling in Section \ref{section:3-1}, i.e., firstly, predefine the similarity threshold \(\alpha\); secondly, classify DD domain fading coefficients using the DD-TCC; finally, continuous time instants that uniformly satisfy the threshold condition are merged to form a complete quasi-invariant interval $T_{QI}$.

In summary, based on the methods described in this section, the complete process of DD domain channel modeling can be summarized as following steps: First, the quasi-stationary interval \(T_{QS}\) is calculated using (\ref{CDD}) for CDD. Then, power distribution fitting is carried out, and the KS test is utilized to select the most suitable distribution. Finally, the quasi-invariant interval \(T_{QI}\) is computed using (\ref{DD-TCC}) for DD-TCC.

\begin{figure*}[ht]
	\centering
	\begin{minipage}[b]{0.3\textwidth}
		\centering
		\includegraphics[width=\textwidth]{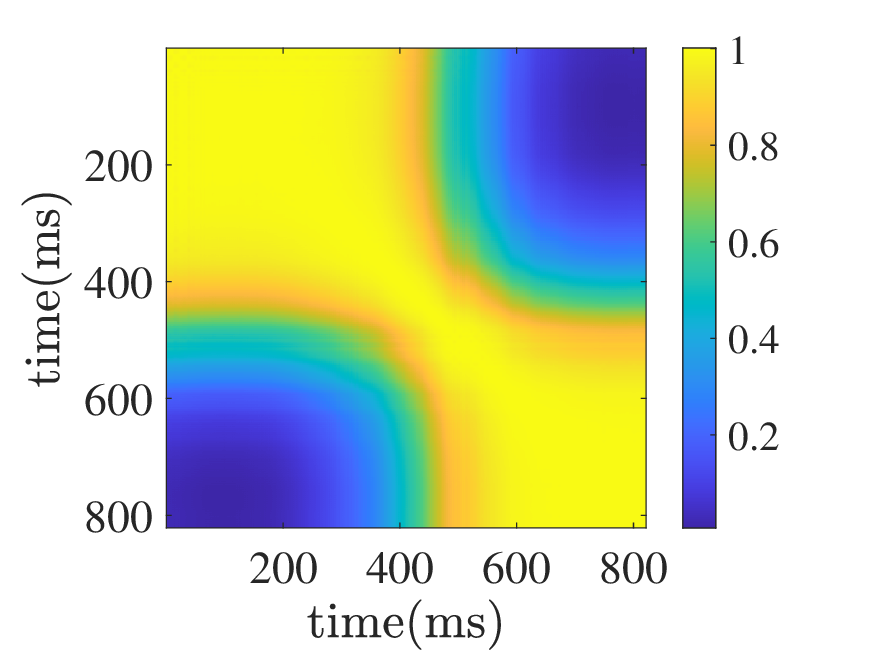}
		\\ \scriptsize(a) Collinearity of DD Domain % 手动添加编号
		\label{fig:Delay-Doppler Domain Characteristic Analysis-1}
	\end{minipage}
	\hfill
	\begin{minipage}[b]{0.3\textwidth}
		\centering
		\includegraphics[width=\textwidth]{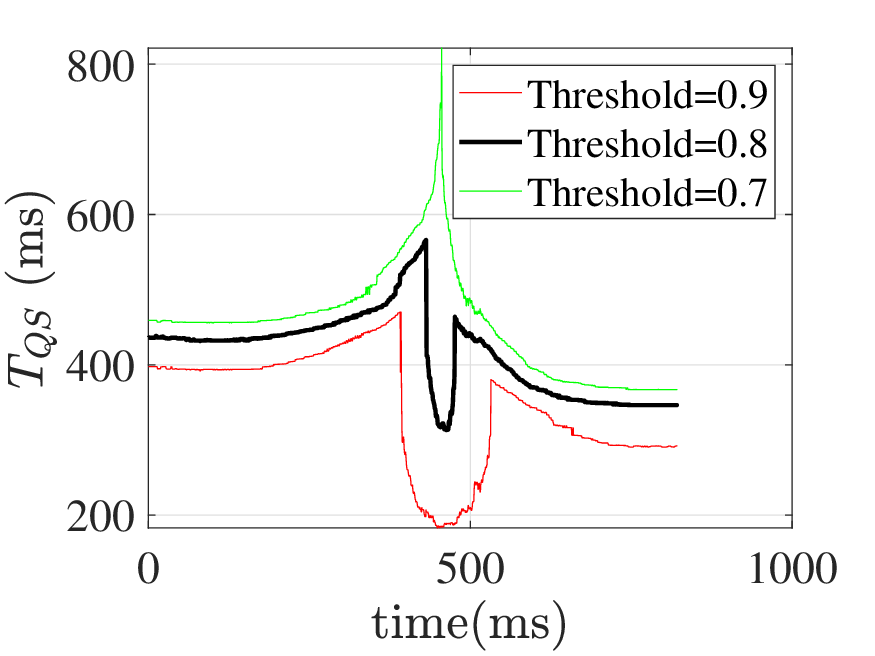}
		\\ \scriptsize(b) CDD Statistics Stationary Interval
		\label{fig:Delay-Doppler Domain Characteristic Analysis-2}
	\end{minipage}
	\hfill
	\begin{minipage}[b]{0.3\textwidth}
		\centering
		\includegraphics[width=\textwidth]{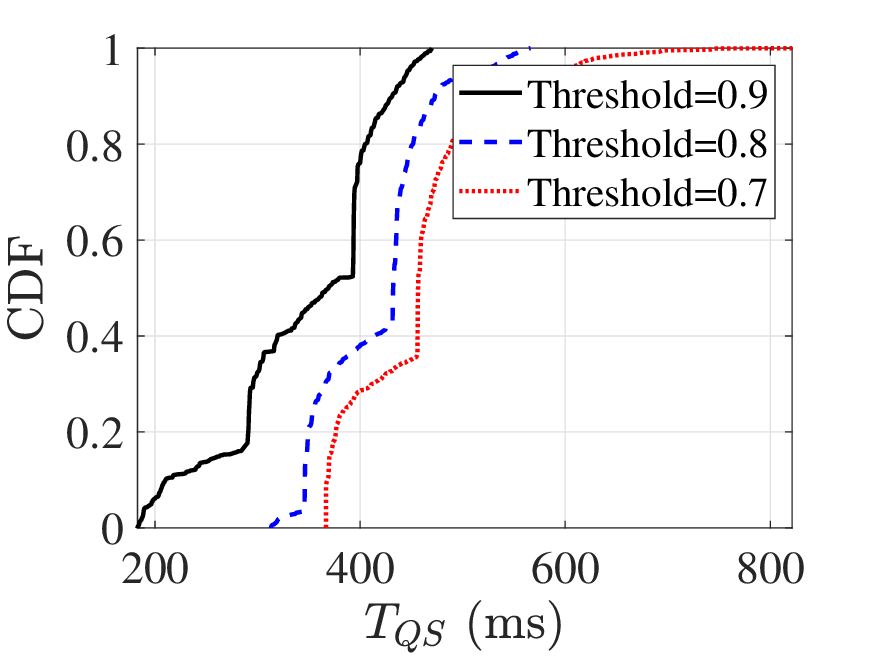}
		\\ \scriptsize(c) CDF
		\label{fig:Delay-Doppler Domain Characteristic Analysis-3}
	\end{minipage}
	\caption{DD domain channel stationarity analysis in HSR weak time-varying scenarios.}
	\label{fig:Delay-Doppler Domain Characteristic Analysis}
\end{figure*}

\begin{figure}[t]
	\centering
	\includegraphics[width=0.95\linewidth]{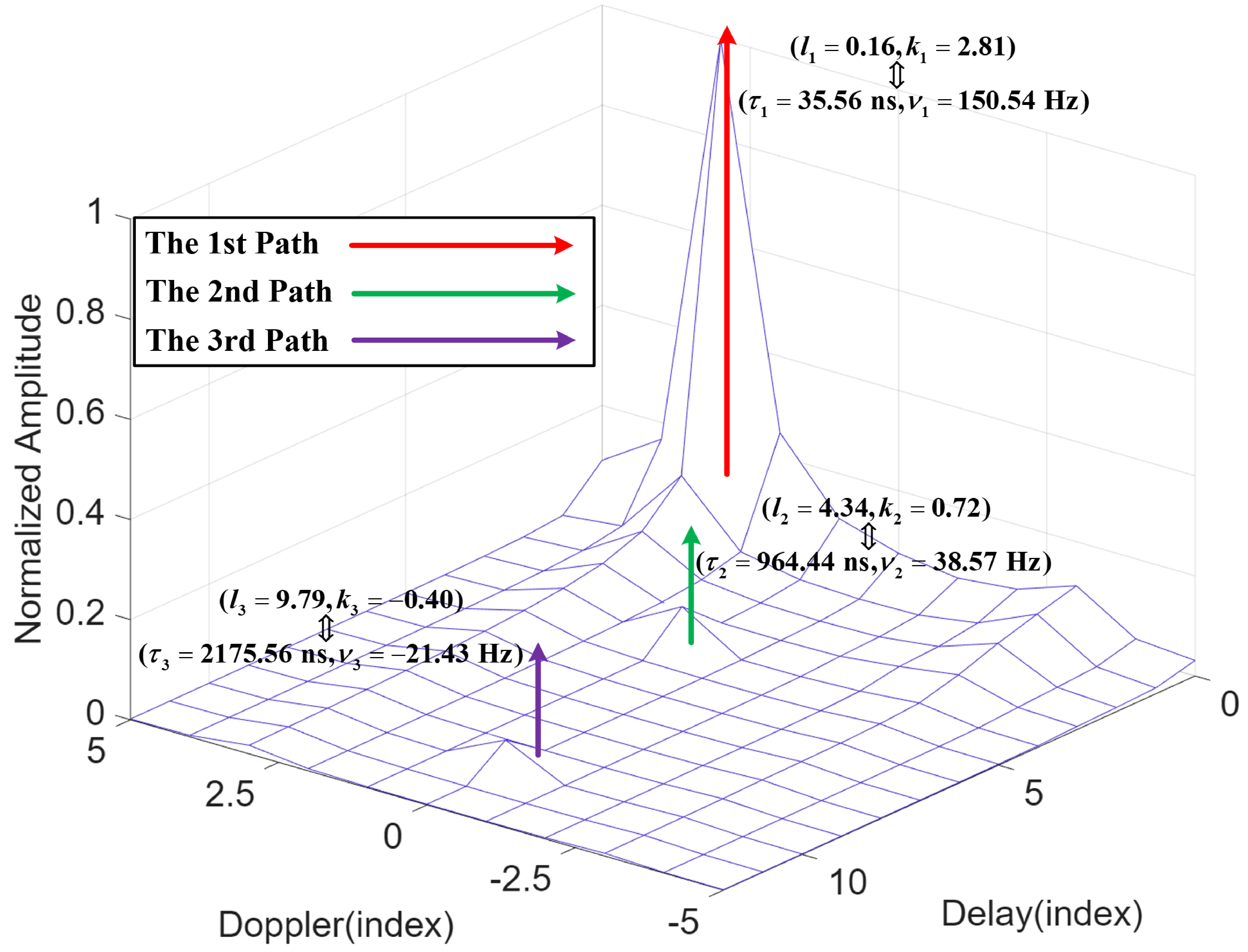}
	\caption{Averaged DD domain channel fading within the stationary intervals in HSR weak time-varying scenarios.} 
	\label{fig:Averaged DD domain channel fading during the station time}
\end{figure}

\begin{table}[t]
	\centering
	\caption{KS Test Fitting Results for Different Probability Distributions in HSR Weak Time-Varying Scenarios}
	\label{tab4:ks_test_results}
	\renewcommand{\arraystretch}{1.5}
	\begin{tabular}{cccc}
		\hline
		\textbf{Number} & \textbf{Distribution} & \textbf{Fitted Parameters} & \textbf{K-S Statistics} \\
		\hline
		\multirow{4}{*}{\textbf{1}} & Rician & s = 0.032, $\sigma$ = 0.004 & 0.0861 \\
		
		& Rayleigh & b = 0.023 & 0.4518 \\
		
		& Nakagami & $\mu$ = 16.703, $\omega$ = 0.001 & 0.1633 \\
		
		& Weibull & a = 0.034, b = 7.383 & 0.2110 \\
		\hline
		\multirow{4}{*}{\textbf{2}} & Rician & s = 0.0003, $\sigma$ = 0.0034 & 0.1442 \\
		
		& Rayleigh & b = 0.0045 & 0.0934 \\
		
		& Nakagami & $\mu$ = 1.4462, $\omega$ = 0 & 0.1432 \\
		
		& Weibull & a = 0.0065, b = 2.033 & 0.1376 \\
		\hline
		\multirow{4}{*}{\textbf{3}} & Rician & s = 0.0001, $\sigma$ = 0.0061 & 0.4623 \\
		
		& Rayleigh & b = 0.0061 & 0.4082 \\
		
		& Nakagami & $\mu$ = 0.4052, $\omega$ = 0.0001 & 0.2575 \\
		
		& Weibull & a = 0.0054, b = 1.1877 & 0.1015 \\
		\hline			
	\end{tabular}
\end{table}

\section{DD Domain Channel Modeling in the HSR Viaduct Scenario}
\label{section:4}
Based on the DD domain channel measurement and modeling method proposed above, this section focuses on establishing DD domain channel models for HSR based on different stationary interval features, aiming to provide a theoretical basis for DDMC communication system design in HSR environments.

The channel data used in this section were collected from a field measurement campaign on the Beijing-Shenyang high-speed railway, covering HSR viaduct scenarios, elevated railway sections with a height of 10-20 m above the ground, surrounded by farmland or low-rise buildings, as described in Section \ref{section:2}. Based on the measured data, this paper classifies the time-varying characteristics of viaduct channels into three intensity levels, corresponding to different stationary intervals characteristics:

(1) Weak time-varying scenarios: In this scenario, the channel is mainly composed of stable line-of-sight (LoS) paths and a small number of fixed reflection paths. The distribution characteristics of the Doppler spectrum do not shift significantly over a long time, and the spectral peak positions and power attenuation laws remain consistent; the survival period of multipath components is long, and the fluctuations in the number of paths and amplitude distribution are extremely small. The quasi-stationary intervals of this scenario is relatively long.

(2) Moderate time-varying scenarios: The overall shape of the Doppler spectrum remains stable within a medium time scale, but some weak multipath components have dynamic fluctuations, and the spectral peak power shows a slow attenuation trend over time; the multipath structure is dominated by stable main paths, and a small number of secondary paths have intermittent changes.

(3) Strong time-varying scenarios: In such scenarios, the shape of the Doppler power spectrum changes significantly within a short time, and the appearance and disappearance of spectral peaks are random; the survival period of multipath components is short, the power attenuation speed is fast, new paths are continuously generated, and old paths disappear quickly. The quasi-stationary intervals of this scenario is relatively short.

The relative positional relationship of the strong, moderate, and weak time-varying scenarios is illustrated in Fig. \ref{fig:Relative Position}, this is similar to the HSR channel characteristics in \cite{ActiveMeasure3-Modeling1}. Based on the classification of the above three time-varying intensity scenarios, this section will carry out targeted DD domain channel modeling work: for weak, moderate, and strong time-varying scenarios, different time windows are adopted to extract channel features, and finally a DD domain channel model that covers the time-varying characteristics of the entire viaduct scenario is formed.

\begin{figure*}[t]
	\centering
	\begin{minipage}[b]{0.3\textwidth}
		\centering
		\includegraphics[width=\textwidth]{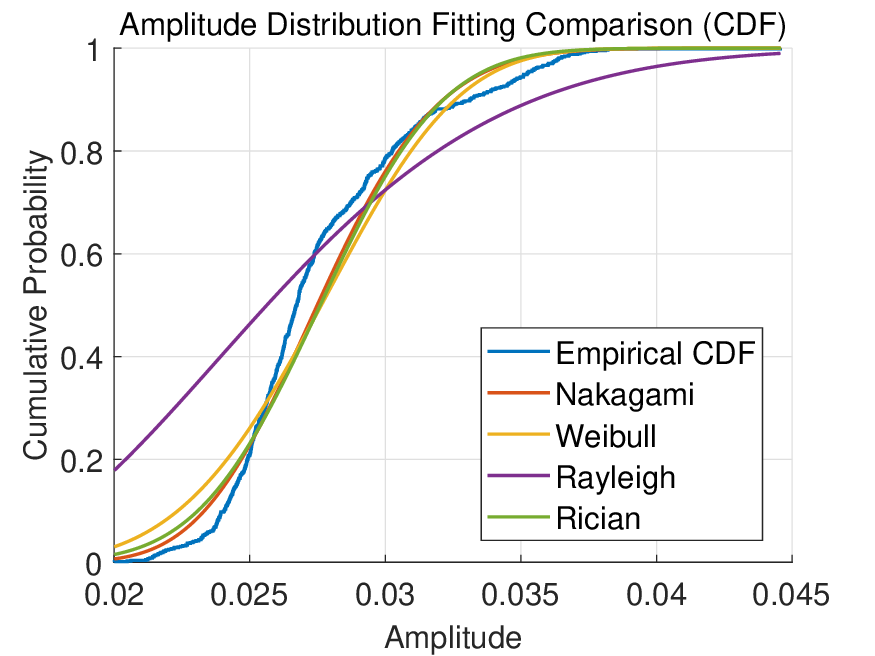}
		\\ \scriptsize(a) The first path 
		\label{fig:CDF Distribution Fitting-1}
	\end{minipage}
	\hfill
	\begin{minipage}[b]{0.3\textwidth}
		\centering
		\includegraphics[width=\textwidth]{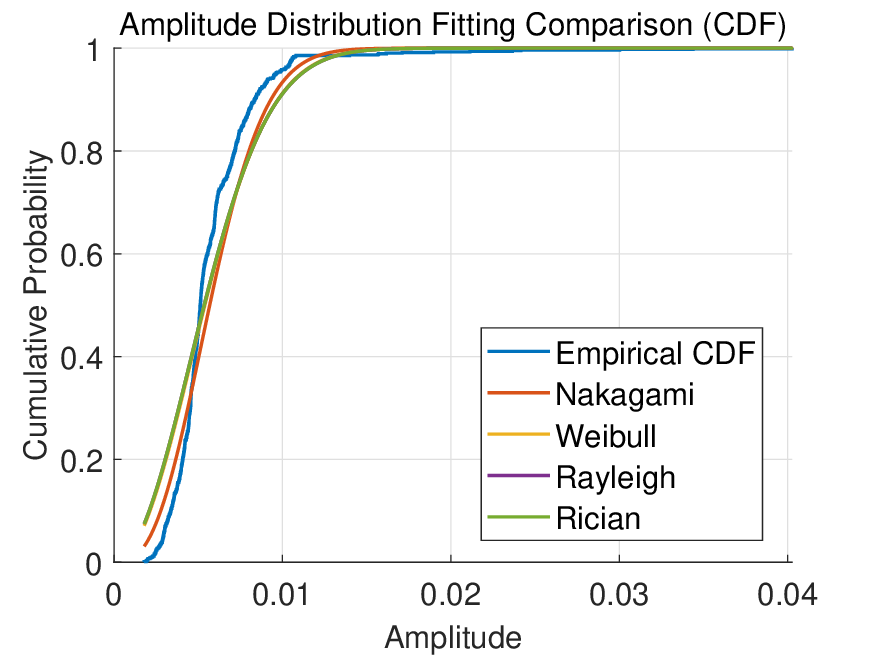}
		\\ \scriptsize(b) The second path
		\label{fig:CDF Distribution Fitting-2}
	\end{minipage}
	\hfill
	\begin{minipage}[b]{0.3\textwidth}
		\centering
		\includegraphics[width=\textwidth]{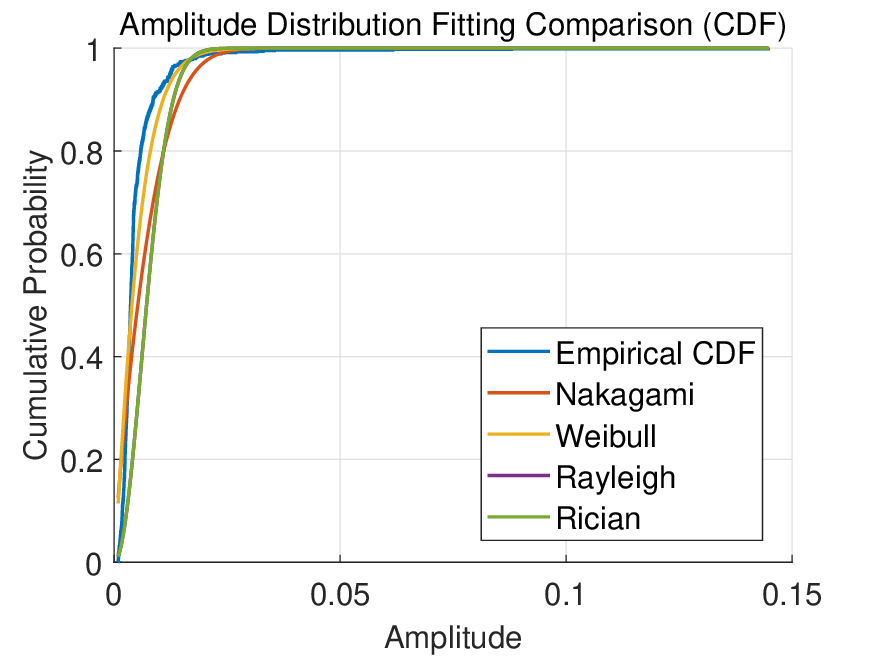}
		\\ \scriptsize(c) The third path
		\label{fig:CDF Distribution Fitting-3}
	\end{minipage}
	\hfill
	\caption{CDF distribution fitting in HSR weak time-varying scenarios.}
	\label{fig:CDF Distribution Fitting}
\end{figure*}

\begin{figure}[t]
	\centering
	\includegraphics[width=0.9\linewidth]{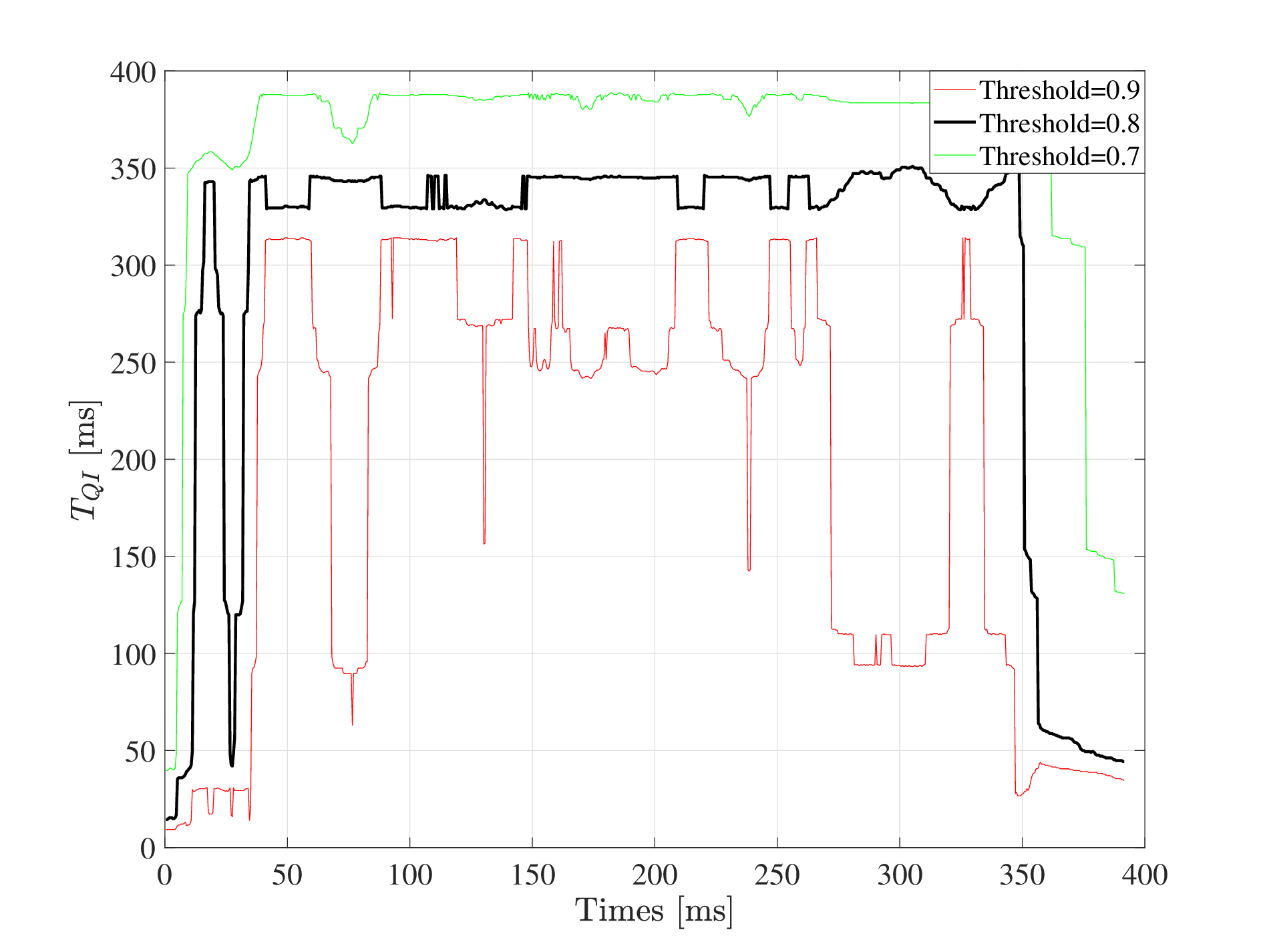}
	\caption{The quasi-invariant interval of the first path in HSR weak time-varying scenarios.} 
	\label{fig:variation within the stationary interval}
\end{figure}

\subsection{Weak Time-Varying Scenarios}
\label{section:4-1}

Taking the weak time-varying scenario as an example, this section outlines the methods and procedures employed in this paper for DD domain channel modeling. Based on the collection of high-speed railway viaduct scene data, firstly, we will extract channel parameter according to Section \ref{section:2-2} and Section \ref{section:2-3}. Secondly, we will analyze the quasi-stationary interval according to Section \ref{section:3-1}, and the calculation results are shown in Fig. \ref{fig:Delay-Doppler Domain Characteristic Analysis}. According to Fig. \ref{fig:Delay-Doppler Domain Characteristic Analysis}, the channel's stationary interval is approximately \( T_{QS} = 392 \) ms when the threshold is greater than 0.9. This means that within this time length, the main statistical characteristics of delay, Doppler, and power of multipath can be considered constant. 

\begin{table}[t]
	\centering
	\caption{Quasi-Invariant Interval Analysis under Different Thresholds in HSR Weak Time-Varying Scenarios}
	\renewcommand{\arraystretch}{1.2}
	\begin{tabular}{cccc}
		\hline
		\textbf{Threshold $\alpha$} & \textbf{$T^{max}_{QI}$ (ms)} & \textbf{$T^{min}_{QI}$ (ms)} & \textbf{$T^{mean}_{QI}$ (ms)} \\
		\hline
		0.9 & 314.07 & 9.33 & 191.88 \\
		
		0.8 & 357.47 & 14.47 & 294.93 \\
		
		0.7 & 388.73 & 39.67 & 363.65 \\
		\hline
	\end{tabular}
	\label{tab5:threshold_analysis}
\end{table}

Thirdly, we model the multipath delay, Doppler, and amplitude behaviors in the quasi-stationary interval according to Section \ref{section:2-4}, as shown in Fig. \ref{fig:Averaged DD domain channel fading during the station time} and Table \ref{tab4-3:DD Channel Model For HSR Under A Viaduct}. From the extracted multipath information, there are three prominent multipath components observed in this test scenario. The first path (marked with a red arrow), characterized by a near-zero delay index and the highest normalized amplitude, is identified as the LoS signal. The Doppler shift of the main path is 150.54 Hz, at this time, the train is far away from the base station, and the Doppler shift of the LoS path is approximately same as the theoretical maximum Doppler shift. For the second multipath, the changes in delay and power indicate that the transmission path with signal attenuation is slightly longer, possibly a path that the signal reaches the receiving end after being reflected by the HSR elevated structures. For the third multipath, we consider the measurement geometry where the BS antenna is installed at a height of about 40 m, while the train operates on a viaduct with a height of 10–20 m and the receiver antenna is mounted 4.05 m above the train roof \cite{PassiveMeasure4}. Due to the significant height difference, the signal can propagate to the ground level below the viaduct. Therefore, this multipath component, which exhibits a larger delay and significant power attenuation, is attributed to scattering from the ground terrain (e.g., farmland or surface clutter) situated below the viaduct level, resulting in a longer propagation path compared to the LoS component.

%For the third multipath, may be the signal is scattered by low lying vegetation, simple sheds, etc, in the farmland around the railway, lead to a further increase in delay and a greater decrease in power, indicating a complex propagation trajectory, severe signal degradation, and a relatively small frequency shift.

Fourthly, we conduct amplitude distribution fitting within the stationary interval according to Section \ref{section:3-2}, and the fitting results are shown in the Fig. \ref{fig:CDF Distribution Fitting} and Table \ref{tab4:ks_test_results}. Therein, the sum of the powers of the multipaths in the DD domain was normalized. Based on the above results, we can observe that: for the first multipath component, among the four distributions, it best fits the Rician distribution, with fitting parameters \( s = 0.032 \) and \( \sigma = 0.004 \), and the K-factor can be obtained by $31.48$ dB. For the second multipath component, the most suitable distribution is the Rayleigh distribution, with fitting parameters \( b = 0.0045 \). The third multipath component follows the Weibull distribution, with fitting parameters \( a = 0.0054 \) and \( b = 1.1877 \).

\begin{figure}[t]
	\centering
	\includegraphics[width=0.88\linewidth]{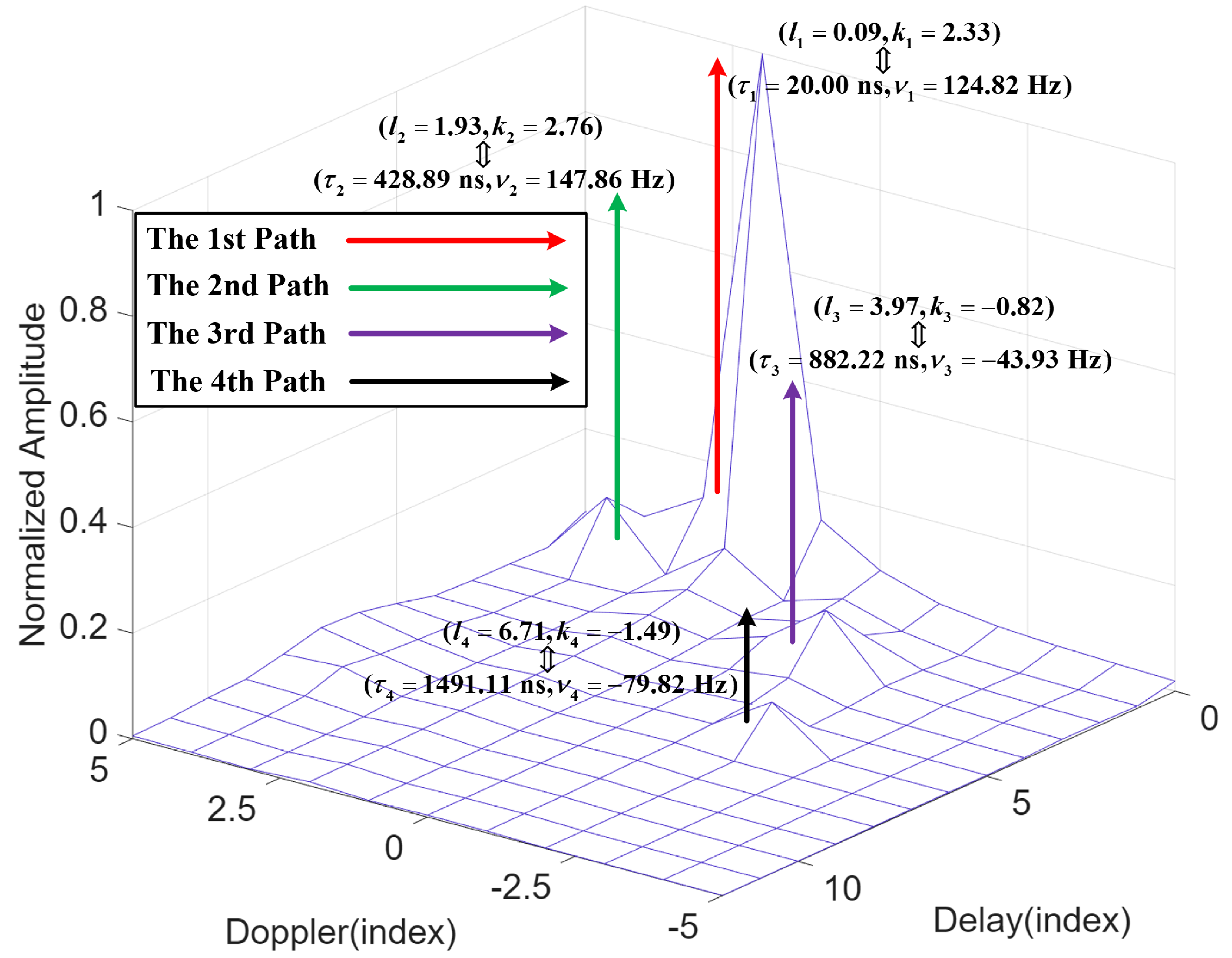}
	\caption{Averaged DD domain channel fading in moderate time-varying scenarios.}
	\label{fig4-2-2:Averaged DD domain channel fading during the station time}
\end{figure}

Finally, we will analyze the quasi-invariant intervals according to Section \ref{section:3-3}. Taking the first path as an example, its quasi-invariant interval within the stationary interval is shown in the Fig. \ref{fig:variation within the stationary interval}. To delve into the impact of different threshold conditions on the quasi-invariant interval, we extract and calculate the maximum value, minimum value, and average value of the quasi-invariant intervals under different thresholds. The results are presented in Table \ref{tab5:threshold_analysis}.

%\begin{figure}[t]
%	\centering
%	\includegraphics[width=\linewidth]{Quasi-Invariant_Interval_at_Different_Thresholds-1.eps}
%	\caption{The quasi-invariant interval at different thresholds in HSR weak time-varying scenarios.} 
%	\label{fig:Stationarity time at different thresholds}
%\end{figure}

Table \ref{tab5:threshold_analysis} reveals distinct patterns in the quasi-invariant interval across the three thresholds, \(T^{max}_{QI}\), \(T^{min}_{QI}\), and \(T^{mean}_{QI}\) denote the maximum, minimum, and average quasi-invariant intervals under different thresholds, respectively. As the threshold decreases from 0.9 to 0.7, a consistent upward trend is observed in all three key indicators. The \(T^{max}_{QI}\) shows a significant increase, rising from $317.07$ ms at the 0.9 threshold to $388.73$ ms at the 0.7 threshold, nearly doubling over the range of threshold values. Similarly, the \(T^{min}_{QI}\) exhibits growth, increasing from $9.33$ ms to $39.67$ ms, indicating a more stable lower bound of quasi-invariant intervals as the threshold decreases. The \(T^{mean}_{QI}\) value, which reflects the overall quasi-invariant performance, also demonstrates a substantial upward trajectory, climbing from $191.88$ ms to $363.65$ ms, highlighting that lower threshold settings are associated with longer average quasi-invariant interval. These trends collectively suggest a clear correlation between threshold levels and the quasi-invariant interval characteristics. 

It is worth noting that in Fig. \ref{fig:variation within the stationary interval}, the fluctuation pattern of the quasi-invariant interval curve at the threshold of $\alpha=0.8$ exhibits distinct behavior compared to those at $\alpha=0.9$ and $\alpha=0.7$. This phenomenon can be attributed to the fact that the threshold of 0.8 lies within the sensitive transition region of the channel's autocorrelation function in this high K-factor scenario. Unlike $\alpha=0.9$, which captures the highly stable LoS core, and $\alpha=0.7$, which encompasses the broader energy envelope, the threshold of 0.8 coincides with the steepest gradient of the correlation decay. In this region, the metric becomes highly sensitive to the transient phase interference patterns between the dominant LoS path and secondary multipath components. Consequently, slight phase rotations can cause the correlation coefficient to oscillate across the threshold, resulting in fluctuations that differ significantly from the trends observed at tighter or looser thresholds.

\subsection{Strong Time-Varying Scenarios}
\label{section:4-3}
\begin{figure}[t]
	\centering
	\includegraphics[width=0.9\linewidth]{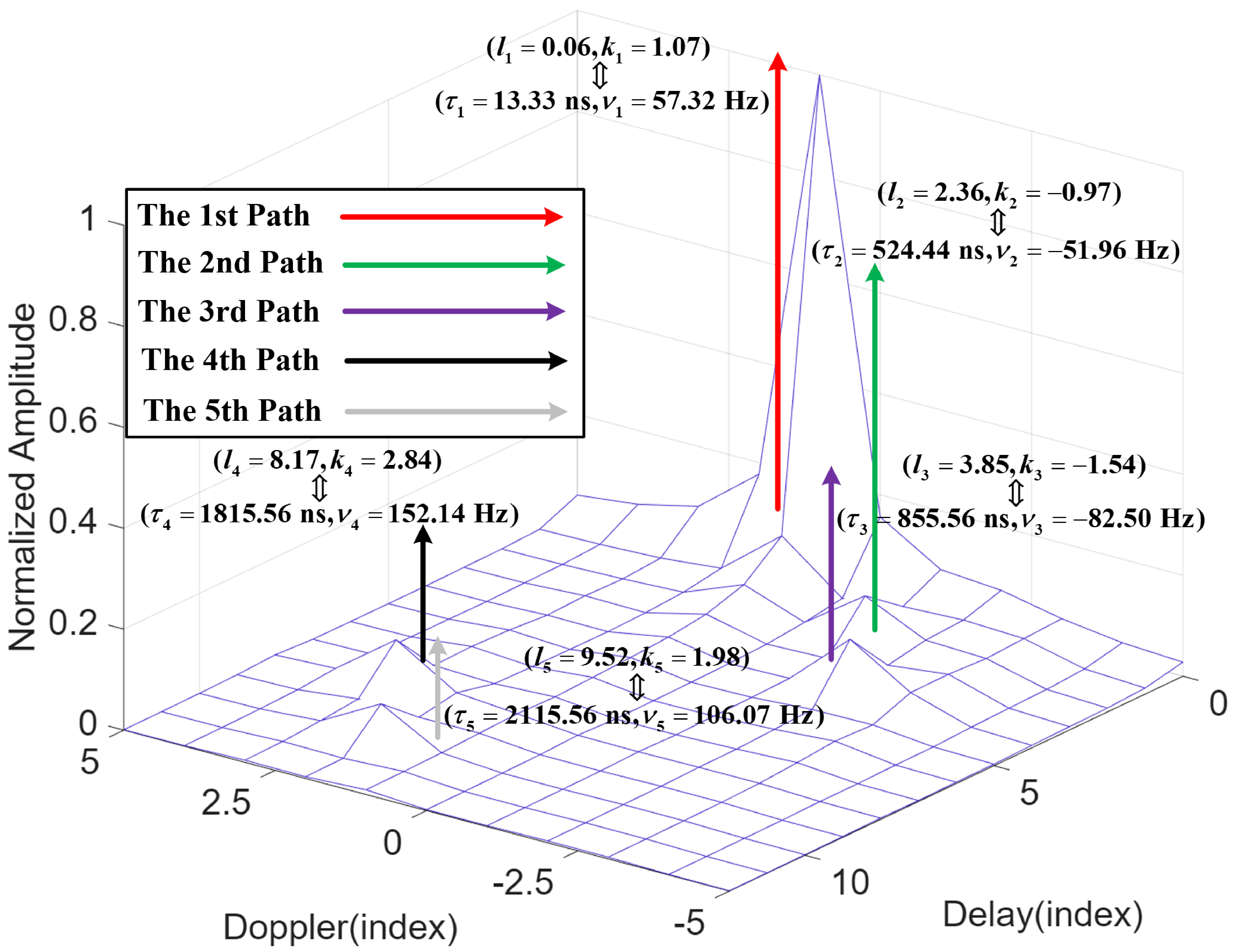}
	\caption{Averaged DD domain channel fading in strong time-varying scenarios.}
	\label{fig4-2-1:Averaged DD domain channel fading during the station time}
\end{figure}

Based on the analysis of the weak time-varying scenario, its key features are as follows: it has a stationary interval $T_{QS}$ of $392$ ms, the average channel fading shows power concentrated in the main path, with a few secondary paths of clear boundaries, and according to Table \ref{tab4-3:DD Channel Model For HSR Under A Viaduct}, it contains 3 multipath components, with LOS propagation dominating.

\subsection{Moderate Time-Varying Scenarios}
\label{section:4-2}

For moderate time-varying scenarios, the multipath structure is relatively stable but still exhibits dynamic fluctuations. Using the quasi-stationary interval evaluation method in Section \ref{section:3-1}, the quasi-stationary interval $T_{QS}$ of this scenario is $172.2$ ms when the threshold is 0.9. The average channel coefficients within this $T_{QS}$ are shown in Fig. \ref{fig4-2-2:Averaged DD domain channel fading during the station time}.

Similar to the weak time-varying case, the first path with the minimum delay corresponds to the LoS component. Compared with weak time-varying scenarios, the amplitude distribution of multipaths in Fig. \ref{fig4-2-2:Averaged DD domain channel fading during the station time} is more dispersed, with an increased number of secondary paths and higher edge ambiguity, indicating an enhancement in multipath dynamics. Using the same channel estimation and multipath extraction methods as applied to weak time-varying scenarios, the channel model parameters for moderate time-varying scenarios are obtained as shown in Table \ref{tab4-3:DD Channel Model For HSR Under A Viaduct}. Table \ref{tab4-3:DD Channel Model For HSR Under A Viaduct} shows that the moderate time-varying scenario contains $4$ main multipath components. The range of multipath Doppler frequency offsets is expanded from $-79.82$ Hz to $147.86$ Hz, reflecting the law that Doppler spread increases as the time-varying intensity of the scenario strengthen. In terms of delay distribution, the maximum delay difference is $1.47$ $\mu$$s$.

\begin{table*}[t]
	\centering
	\caption{DD Channel Model For HSR Under A Viaduct}
	\label{tab4-3:DD Channel Model For HSR Under A Viaduct}
	\renewcommand{\arraystretch}{1.8}
	\begin{tabular}{cccccccc}
		\hline
		\textbf{\makecell{Name \\ (Scenarios)}} & \textbf{\makecell{Stationary Interval \\ $T_{QS}$}} & \textbf{Number} & \textbf{Delay(ns)} & \textbf{Power(dB)} & \textbf{Doppler(Hz)} & \textbf{Amplitude Distribution} & \textbf{\makecell{$T^{min}_{QI}$ \\ ($\alpha = 0.9$)}}\\
		\hline
		\multirow{5}{*}{\textbf{\makecell{TDDL-A\\(Strong\\Time-Varying)}}} & \multirow{5}{*}{100.8 ms} & 1 & 0 & 0 & 57.32  & Rician$\sim$(0.063,0.014), $K$=11.04 dB & 2.8 ms \\
		&
		
		& 2 & 511.11 & -26.33  & -51.96  & Rician$\sim$(0.008,0.001), $K$=25.89 dB & 1.87 ms \\&
		
		& 3 & 842.23 & -23.32  & -82.50  & Rician$\sim$(0.007,0.001), $K$=13.33 dB & 1.4 ms \\&
		
		& 4 & 1802.23 & -26.16  & 152.14  & Rician$\sim$(0.006,0.002), $K$=6.45 dB & 1.4 ms \\&
		
		& 5 & 2102.23 & -28.57  & 106.07  & Weibull$\sim$(0.008,1.38) & 0.93 ms \\
		\hline
		\multirow{4}{*}{\textbf{\makecell{TDDL-B \\ (Moderate\\Time-Varying)}}} & \multirow{4}{*}{172.2 ms} & 1 & 0 & 0 & 124.82 & Rician$\sim$(0.044,0.016), $K$=3.81 dB & 4.67 ms \\&
		
		& 2 & 408.89 & -20.99 & 147.86 & Weibull$\sim$(0.009,1.517) & 1.87 ms \\&
		
		& 3 & 862.22 & -25.34  & -43.93  & Weibull$\sim$(0.007,1.294) & 0.93 ms \\&
		
		& 4 & 1471.11 & -27.18  & -79.82  & Weibull$\sim$(0.005,1.353) & 0.93 ms \\
		\hline
		\multirow{4}{*}{\textbf{\makecell{TDDL-C \\ (Weak\\Time-Varying)}}} & \multirow{3}{*}{392 ms} & 1 & 0 & 0 & 150.54 & Rician$\sim$(0.032,0.004), $K$=31.48 dB & 9.33 ms \\&
		
		& 2 & 928.88 & -28.74  & 38.57 & Rayleigh$\sim$(0.0045) & 1.87 ms \\&
		
		& 3 & 2140 & -30.45  & -21.43  & Weibull$\sim$(0.0054,1.1877) & 0.93 ms \\
		\hline			
	\end{tabular}
\end{table*}

Considering the characteristics of high multipath dynamics and severe Doppler spectrum fluctuations in viaduct scenarios. Using the quasi-stationary interval evaluation method in Section \ref{section:3-1}, the quasi-stationary interval $T_{QS}$ of this scenario is $100.8$ ms when the threshold is 0.9.

Based on the DD domain channel modeling framework proposed in Section \ref{section:3}, the average channel coefficients within this stationary interval are obtained in Fig. \ref{fig4-2-1:Averaged DD domain channel fading during the station time}. As indicated by the red arrow, the first path represents the LoS signal, serving as the dominant propagation path. Fig. \ref{fig4-2-1:Averaged DD domain channel fading during the station time} clearly presents the power aggregation characteristics in the DD domain: multiple secondary paths are distributed around the main path, which confirms the characteristic of rapid multipath changes in strong time-varying scenarios. 

Through channel estimation methods and multipath extraction methods, channel model parameters for this scenario are obtained as shown in Table \ref{tab4-3:DD Channel Model For HSR Under A Viaduct}. It can be seen from the table that there are a total of $5$ significant multipath components in strong time-varying scenario. The Doppler frequency offset of multipaths has a wide distribution range, indicating that the Doppler spread effect caused by high-speed movement is significant. Meanwhile, the maximum multipath delay difference reaches $2.1$ $\mu$$s$, which reflects the multipath propagation characteristics in complex scattering environments.

\subsection{DD Domain Channel Model Analysis}
\label{section:4-4}
Through the modeling and parameter extraction of the three types of time-varying scenarios (strong, moderate, and weak), the DD domain channel models of the HSR viaduct scenario are collected in Table \ref{tab4-3:DD Channel Model For HSR Under A Viaduct}. In line with the tapped delay line (TDL) model \cite{TDL}, we regard this model as the tapped delay Doppler line (TDDL) model and name it TDDL-A, TDDL-B, and TDDL-C respectively.

The results in Table \ref{tab4-3:DD Channel Model For HSR Under A Viaduct} show that in the TDDL-A model ($100.8$ ms stationary interval, $5$ multipaths), delay spans up to $2102.23$ ns, Doppler ranges widely (from $-82.50$ Hz to $152.14$ Hz), and mean quasi-invariant intervals are short (down to $5.81$ ms), reflecting complex scattering and rapid channel changes. For the TDDL-B model ($172.2$ ms stationary interval, 4 multipaths), delay (up to $1471.11$ ns) and Doppler (from $-79.82$ Hz to $147.86$ Hz) ranges narrow, amplitude is mainly Weibull, and mean quasi-invariant intervals are longer (up to $30.12$ ms), indicating a more regular scattering and slower changes. In the TDDL-C model ($392$ ms stationary interval, $3$ multipaths), though delay reaches $2140$ ns, multipath count is low, the main path dominates, Doppler has narrow fluctuations (from $-21.43$ Hz to $150.54$ Hz), amplitude is mostly Rician, and mean quasi-invariant intervals are relatively long (up to $191.88$ ms), showing stable propagation with few random scatterers. From strong to weak time-varying, multipath count drops, stationary interval and quasi-invariant intervals increase, amplitude distribution simplifies, and Doppler/power fluctuations narrow. These characteristics are consistent with real world propagation behaviors, and could support future DDMC and integrated sensing and communication designs for 6G and beyond.

\section{Verification  of  DD Domain Channel Models  in  Viaduct}
\label{section:5}
%In this section, we propose to verify the accuracy of the obtained DD domain channel model by comparing the BER performance of the communication system under the measured channel and the modeled channel. 
In this section, we provide a comprehensive verification of the proposed DD domain channel models through both physical dispersion analysis and system-level performance evaluation. We first validate the physical fidelity of the models by comparing the RMS delay and Doppler spreads with the measured statistics. Subsequently, the BER performance is employed as a statistical utility metric to demonstrate the model's accuracy in capturing the channel's fading characteristics.
\subsection{Verification of Dispersion Characteristics and Modeling Fidelity}
\label{section:5-0}
To provide a verification of the modeling fidelity, we first evaluate the consistency of physical dispersion parameters between the measurement data and the proposed TDDL models. We adopt the RMS delay spread $\tau_{rms}$ and RMS Doppler spread $\nu_{rms}$ as key benchmarks to quantify the channel dispersion, which are defined as \cite{ChannelEstimation1,TDL}:
\begin{equation}
	\tau_{rms} = \sqrt{\frac{\sum P_i \tau_i^2}{\sum P_i} - \left( \frac{\sum P_i \tau_i}{\sum P_i} \right)^2},
\end{equation}
\begin{equation}
	\nu_{rms} = \sqrt{\frac{\sum P_i \nu_i^2}{\sum P_i} - \left( \frac{\sum P_i \nu_i}{\sum P_i} \right)^2},
\end{equation}
where $P_i$, $\tau_i$, and $\nu_i$ denote the power, delay, and Doppler shift of the $i$-th multipath component, respectively.

\begin{table}[h]
	\centering
	\caption{Comparison of RMS Dispersion Parameters}
	\renewcommand{\arraystretch}{1.5}
	\begin{tabular}{cccc}
		\hline
		\textbf{Scenarios} & \textbf{Data Source} & \textbf{\makecell{RMS Delay \\ $\tau_{rms}$ (ns)}} & \textbf{\makecell{RMS Doppler  \\ $\nu_{rms}$ (Hz)}} \\
		\hline
		\multirow{4}{*}{\textbf{TDDL-A}} & Measured within $T_{QS}$ & 135.23 & 12.51 \\&
		
		Modeled in Table V & 132.58 & 11.91 \\&
		
		Measured within $\frac{1}{2}T_{QS}$ & 135.92 & 12.47 \\&
		
		Measured within $2T_{QS}$ & 162.45 & 15.82 \\
		\hline
		\multirow{4}{*}{\textbf{TDDL-B}} & Measured within $T_{QS}$ & 88.99 & 13.20 \\&
		
		Modeled in Table V & 86.51 & 12.83 \\&
		
		Measured within $\frac{1}{2}T_{QS}$ & 86.55 & 12.84 \\&
		
		Measured within $2T_{QS}$ & 76.12 & 9.35 \\
		\hline
		\multirow{4}{*}{\textbf{TDDL-C}} & Measured within $T_{QS}$ & 72.41 & 6.56 \\&
		
		Modeled in Table V & 72.49 & 6.57 \\&
		
		Measured within $\frac{1}{2}T_{QS}$ & 74.06 & 6.71 \\&
		
		Measured within $2T_{QS}$ & 89.33 & 4.15 \\
		\hline
	\end{tabular}
	\label{tabV-A:RMS DD}
\end{table}

\begin{figure*}[t]
	\centering
	\begin{minipage}[b]{0.3\textwidth}
		\centering
		\includegraphics[width=\textwidth]{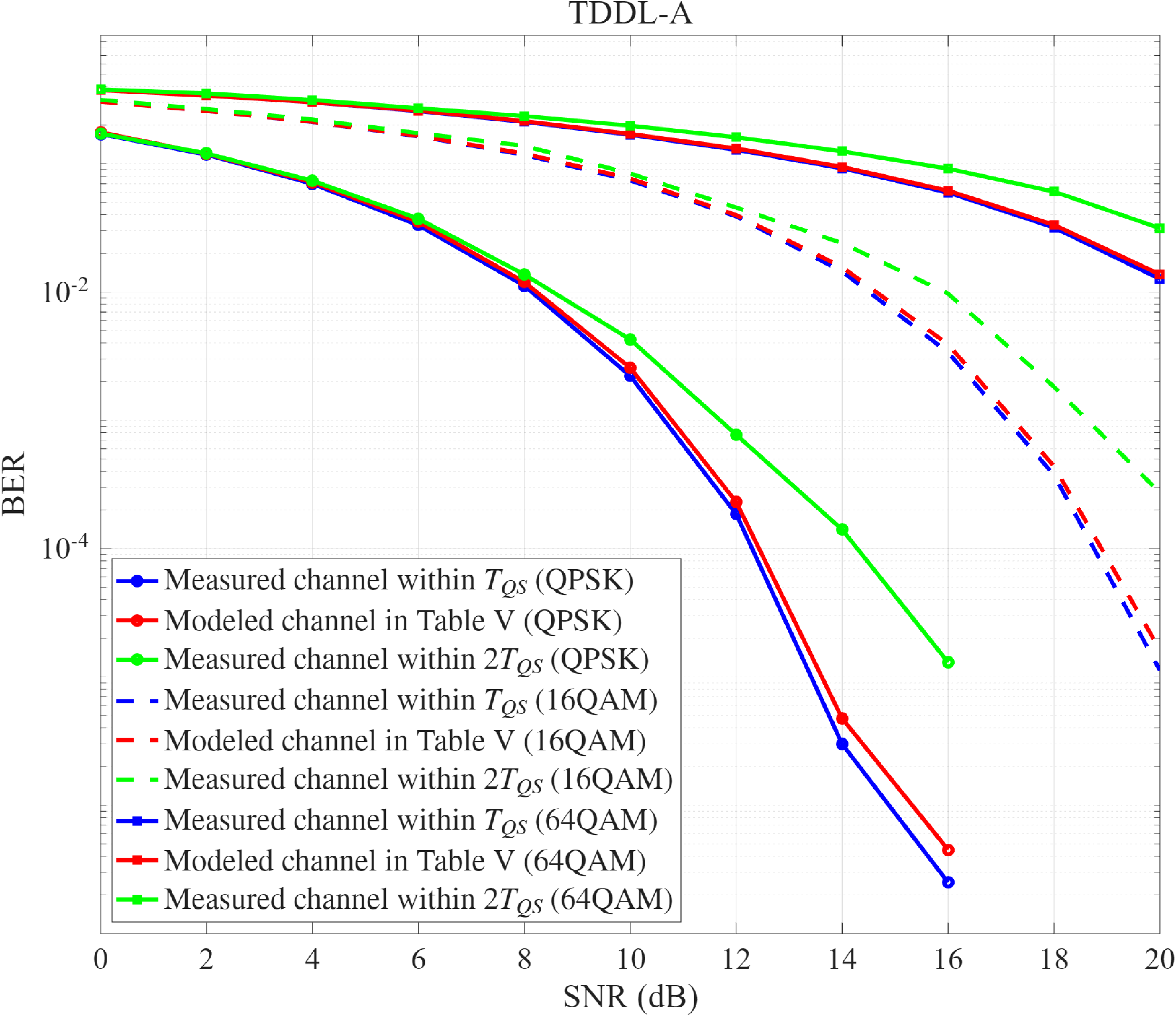}
		\\ \scriptsize(a) TDDL-A 
		\label{fig:TDDL-A}
	\end{minipage}
	\hfill
	\begin{minipage}[b]{0.3\textwidth}
		\centering
		\includegraphics[width=\textwidth]{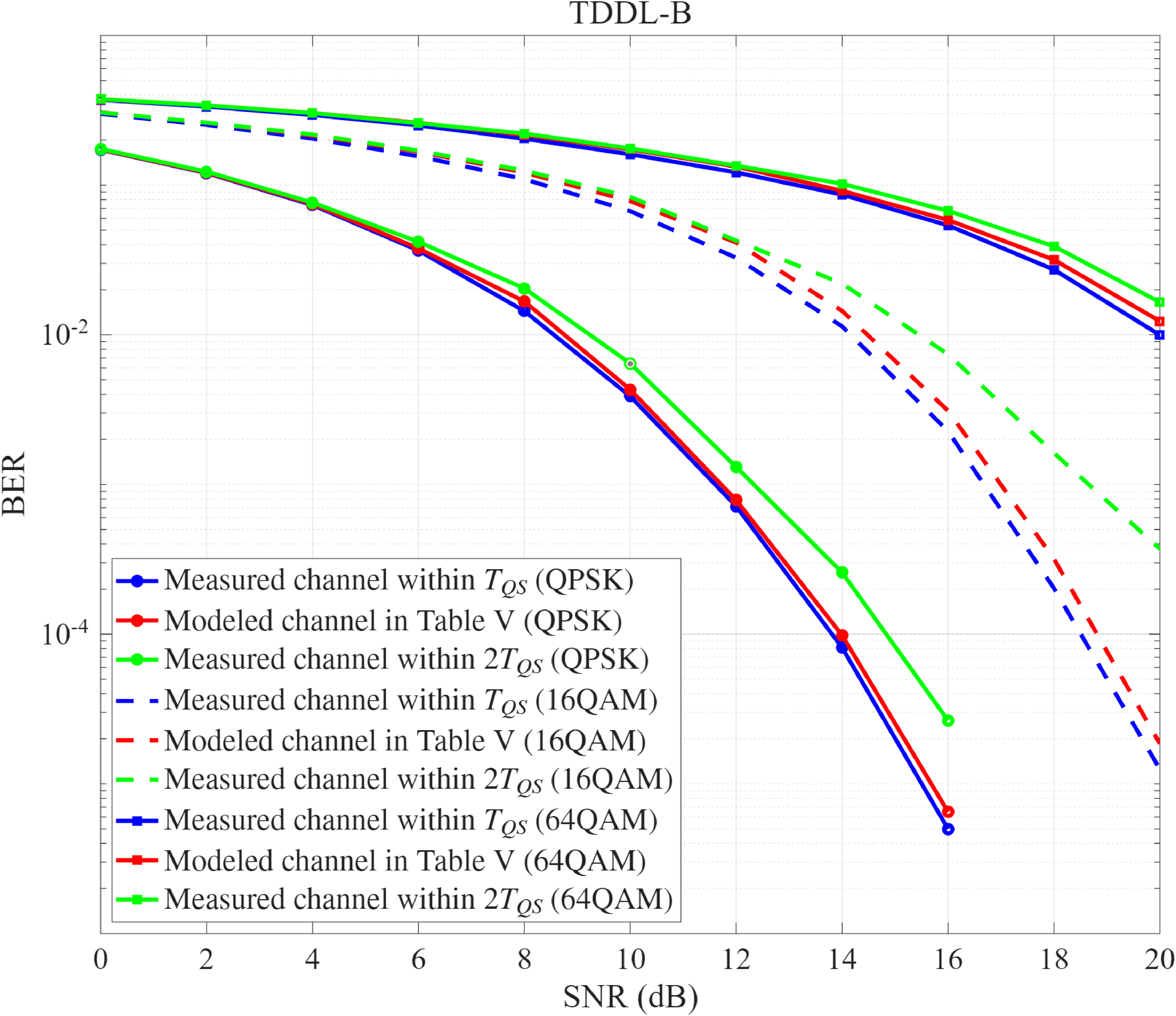}
		\\ \scriptsize(b) TDDL-B
		\label{fig:TDDL-B}
	\end{minipage}
	\hfill
	\begin{minipage}[b]{0.3\textwidth}
		\centering
		\includegraphics[width=\textwidth]{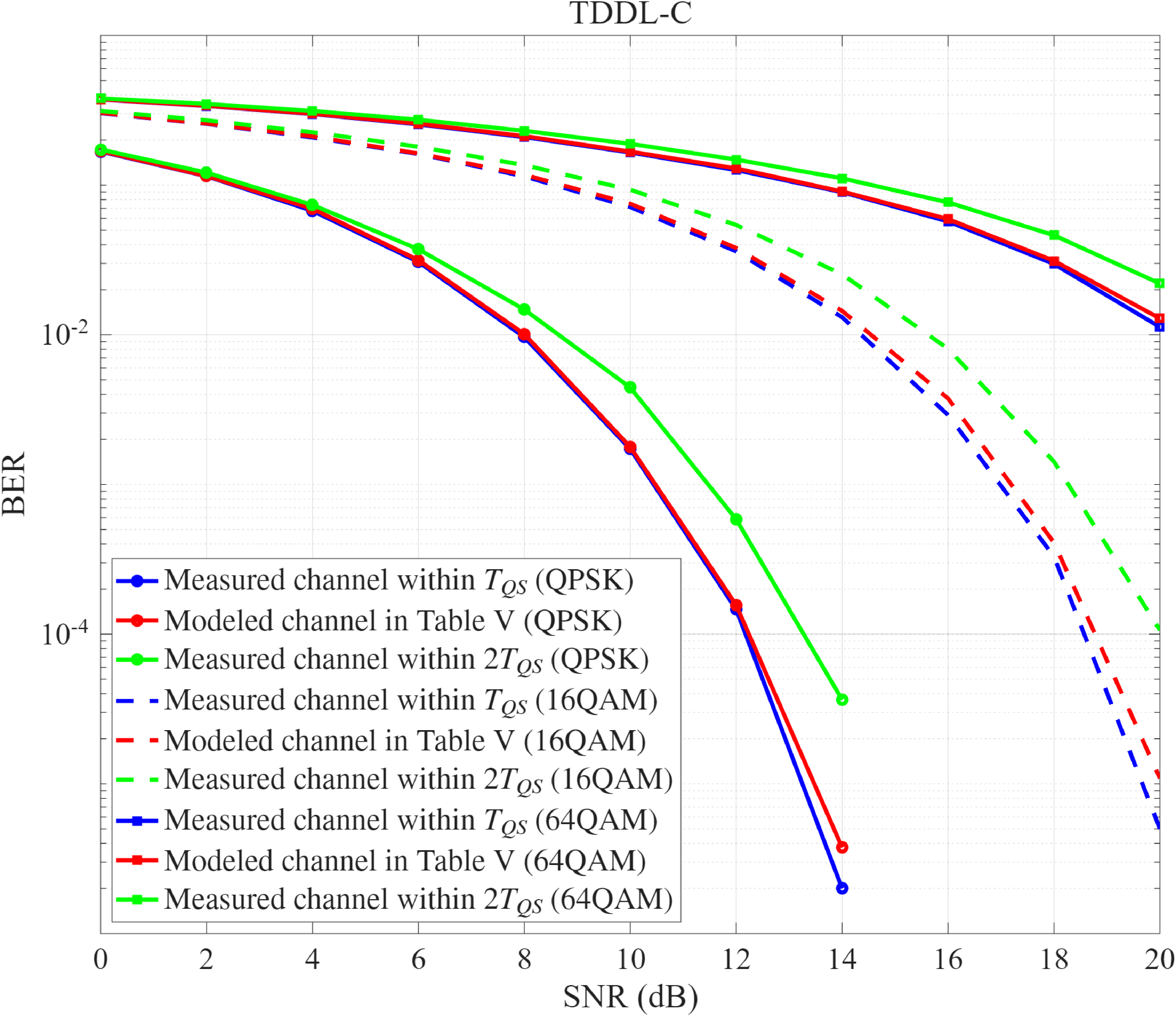}
		\\ \scriptsize(c) TDDL-C
		\label{fig:TDDL-C}
	\end{minipage}
	\hfill
	\caption{BER of OTFS modulation under different channel conditions.}
	\label{fig:BER Comparison Across Different Time-Varying Scenarios}
\end{figure*}

To rigorously verify the accuracy of the established three TDDL models and validate the rationality of the determined quasi-stationary intervals, we performed a comparative analysis using measurement data from three distinct time durations: the identified quasi-stationary interval ($T_{QS}$), half of the interval ($\frac{1}{2}T_{QS}$), and double the interval ($2T_{QS}$). The comparison results are summarized in Table \ref{tabV-A:RMS DD}. 

It can be observed that for all three models, the RMS delay and Doppler spreads of the TDDL models exhibit consistency with the measurement results within $T_{QS}$, confirming the accuracy of the parameter extraction. Furthermore, the statistics derived from the shorter duration ($\frac{1}{2}T_{QS}$) align closely with those of $T_{QS}$, verifying the stability of the channel statistics within the modeled interval. In contrast, when the observation window is extended to twice the stationary interval ($2T_{QS}$), significant deviations in dispersion parameters are observed compared to the model. For example, in the strong time-varying scenario (TDDL-A), the $\tau_{rms}$ increases significantly from 135.23 ns to 162.45 ns. This deviation confirms that the channel characteristics undergo significant changes beyond the defined $T_{QS}$, thereby validating the accuracy of the quasi-stationary interval evaluation presented in Section III.

\subsection{Verification of the Accuracy of Channel Statistical Distributions within the Quasi-Stationary Interval in the DD Domain}
\label{section:5-2}

Using BER as a ‘utility metric’ to verify the accuracy of channel models is the standard paradigm in current 5G/6G high mobility channel modeling research \cite{Off-grid,BER1,Add1}. The BER performance of systems (OFDM/OTFS) depends strictly on the power delay distribution and Doppler power spectrum of the channel, and deviations in these parameters will cause significant shifts in the BER curve \cite{BER2}. Therefore, the BER of the model channel and the measured channel can effectively prove whether the proposed model can accurately capture the dispersion and temporal correlation of the physical channel. Since the BER performance of the DD domain modulated system is closely relevant with the DD domain channel characteristic, we select the OTFS waveform as an example for channel models' verification. Under the ideal waveform assumption, the input-output relationship of the OTFS system is expressed as \cite{Off-grid,ChannelEstimation2}:
\begin{equation}
	\label{y[k,l]}
	\small
	\begin{split}
	y[k,l]& = \frac{1}{NM} \sum_{k'=0}^{N-1} \sum_{l'=0}^{M-1} x[k',l'] \sum_{i=1}^{P} h_i e^{-j2\pi \nu_i \tau_i} \\
	&\sum_{n'=0}^{N-1} e^{\frac{-j2\pi (k - k' - k_i)n'}{N}} \sum_{m'=0}^{M-1} e^{\frac{-j2\pi (l - l' - l_i)m'}{M}} + n[k,l],
	\end{split}
\end{equation}
where \(y[k,l]\) denotes the output signal, \(x[k',l']\) denotes the input signal, \(n[k,l]\) denotes additive noise. Based on (\ref{y[k,l]}), we can obtain:
\begin{equation}
	\label{18}
	{\bf Y}_{\rm DD} = {\bf H}_{\rm DD}{\bf X}_{\rm DD} + {\bf n},
\end{equation}
where ${\bf Y}_{\rm DD} \in \mathbb{C}^{NM \times 1}$ denotes the vectorized received symbols, ${\bf X}_{\rm DD} \in \mathbb{C}^{NM \times 1}$ denotes the vectorized transmitted symbols, and ${\bf H}_{\rm DD} \in \mathbb{C}^{NM \times NM}$ denotes the DD domain equivalent channel matrix. After transmission over the DD domain channel, the received signal can be equalized with the minimum mean square error (MMSE) criterion:
\begin{equation}
	\label{XDD}
	\hat{\bf X}_{\rm DD} = {\bf H}_{\rm DD}^H ({\bf H}_{\rm DD} {\bf H}_{\rm DD}^H + \frac{\sigma^2}{E_{{\bf X}_{\rm DD}}}{\mathbb I}_{NM})^{-1} {\bf Y}_{\rm DD}.
\end{equation}
where \(\sigma^2\) and \(E_{{\bf X}_{\rm DD}}\) denote the noise power and symbol power, respectively. Subsequently, the BER of the system transmission can be obtained by comparing \(\hat{\bf X}_{\text{DD}}\) and \(\bf X_{\text{DD}}\).

Thereafter, based on (\ref{y[k,l]})-(\ref{XDD}), we can compare the BER performances of the OTFS system under the measured DD domain channel and the modeled channel under different signal-to-noise ratio (SNR) conditions, so as to verify the accuracy of the derived channel model. The specific details are described as follows.

To verify the accuracy of the DD domain channel model established in Table \ref{tab4-3:DD Channel Model For HSR Under A Viaduct}, this section compares the BER performance of the OTFS system under the measured channel and the modeled channel within the quasi-stationary interval \(T_{QS}\). Regarding the simulation conditions, we assume ideal time-frequency synchronization and perfect channel state information at the receiver \cite{Ideal1}. It is important to note that in practical deployments, the absolute BER performance would inevitably be degraded by factors such as channel estimation errors, synchronization offsets, carrier frequency offsets, and hardware nonlinearities. However, the primary objective of this simulation is not to benchmark the performance of a practical OTFS transceiver, but rather to utilize BER as a "utility metric" to validate the statistical fidelity of the proposed channel models. By assuming ideal conditions for both the measured and modeled channel simulations, we strictly isolate the channel modeling errors from implementation-dependent factors. Introducing hardware impairments or specific receiver algorithms would introduce nuisance variables that could obscure the intrinsic deviations of the channel model \cite{Ideal2}. Therefore, the BER results presented here serve as a verification of the measurement and modeling accuracy. While the absolute BER values will vary under different OTFS transceiver designs and practical impairment levels, the relative agreement between the model and measurement under these controlled conditions confirms the model's validity. For OTFS transmission under the measurement channel, the DD domain channel corresponding to the duration $NT$ and bandwidth $M\Delta f$ is randomly selected from the measurement DD domain channel with the quasi-stationary interval $T_{QS}$. Besides, for OTFS transmission under the modeled channel, the DD domain channel corresponding to the RE is generated according to parameters and distributions modeled in Table \ref{tab4-3:DD Channel Model For HSR Under A Viaduct}. To verify whether the established DD domain channel models are only valid within the quasi-stationary interval, we select the OTFS transmission with measured DD domain channel selected from a time duration longer than the quasi-stationary interval as an additional benchmark,  e. g., within twice the stationary interval (\(2T_{QS}\)). Based on above settings, BER of different schemes will be calculated based on (\ref{y[k,l]})-(\ref{XDD}). In the following, we set $M=300$ and $N = 280$, and the QPSK, 16QAM and 64QAM modulation is used.

The BER performance of OTFS modulation under different DD domain channel conditions is shown in Fig. \ref{fig:BER Comparison Across Different Time-Varying Scenarios}. To provide a comprehensive validation of the proposed TDDL models' versatility and robustness, we evaluate the system performance using QPSK, 16QAM, and 64QAM modulations. First, it can be observed that for all three modulation schemes, the BER curves of OTFS under the proposed channel models align closely with those under measured channels randomly selected from the corresponding quasi-stationary intervals. Specifically, the tight alignment even at 64QAM demonstrates that the established DD domain channel models accurately capture the fine-grained statistical characteristics and dispersion properties required for high-order modulations. This indicates that in strong, moderate, and weak time-varying scenarios, the proposed models exhibit high fidelity and robustness. Second, when comparing the BER curves under the modeled channels with those under measured channels randomly chosen from $2T_{QS}$ durations, a BER gap emerges due to the introduction of DD domain channel non-stationarity. Based on these observations, we can conclude that the proposed TDDL-A, TDDL-B, and TDDL-C channel models are valid within corresponding quasi-stationary intervals and can serve as references of the realistic channel to simulate DD domain channel for the data package with time duration $NT$ when $NT<T_{QS}$.

\begin{figure}[t]
	\centering
	\includegraphics[width=0.9\linewidth]{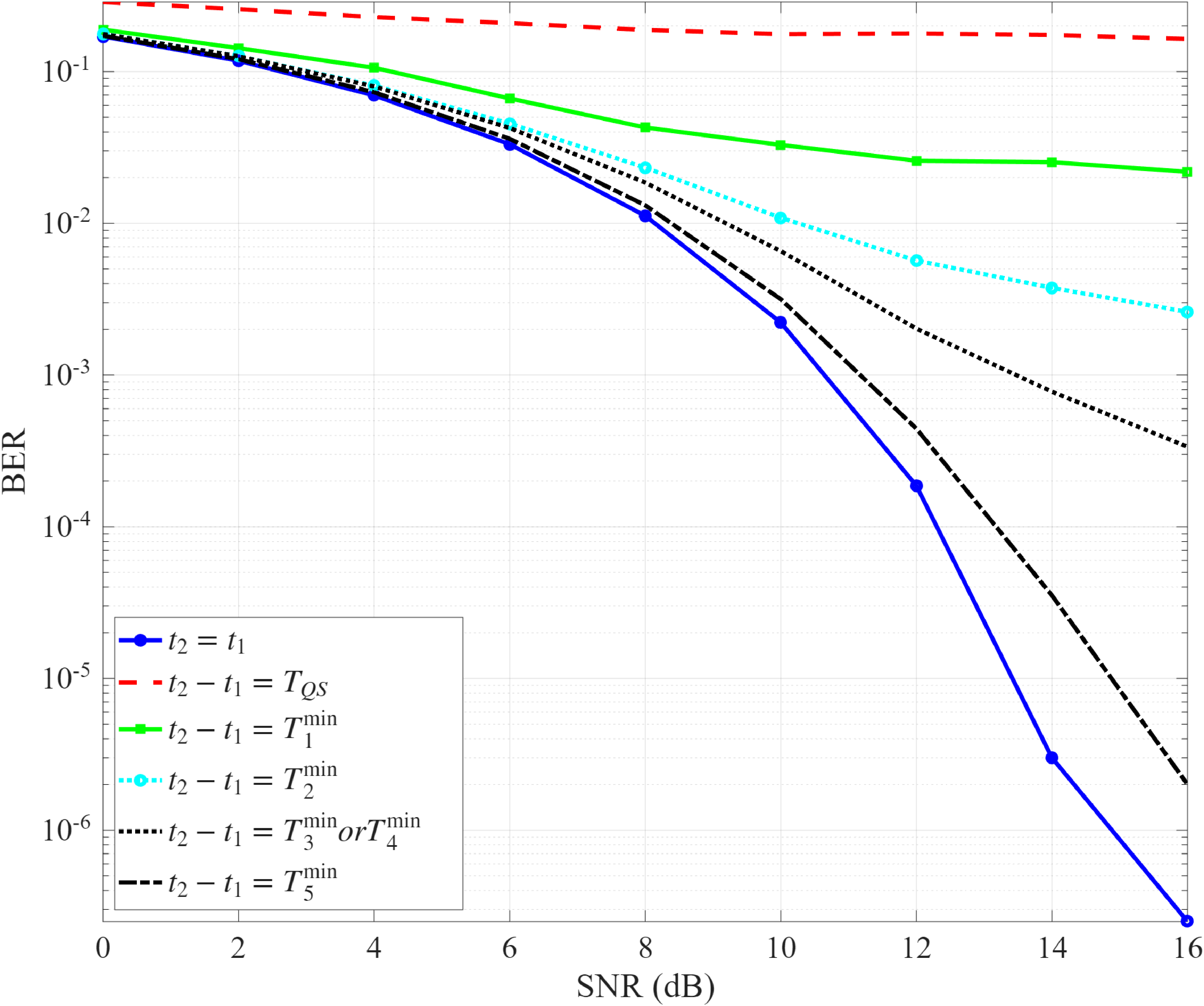}
	\caption{BER of OTFS modulation under different $t_2-t_1$.}
	\label{fig:BER Comparison across TDDL-A}
\end{figure}

\subsection{Verification of Quasi-Invariant Intervals}
\label{section:5-3}

In this subsection, we will further verify that the DD domain channel fading coefficient can be regarded invariant within the quasi-invariant interval. In detail, we will use $\bf H_{\rm DD}$ at instant ${t_2}$ for OTFS transmission in (\ref{18}), and $\bf H_{\rm DD}$ at instant ${t_1}$ for channel equalization in (\ref{XDD}), which are denoted as ${\bf H}^{t_2}_{\rm DD}$ and ${\bf H}^{t_1}_{\rm DD}$ with $t_2 \le t_1$. Besides, ${\bf H}^{t_2}_{\rm DD}$ and ${\bf H}^{t_1}_{\rm DD}$ will be generated according to the measured DD domain channel at corresponding time instants. During the simulation, the BER results of three schemes under the strong time-varying scenario are compared: the first scheme is called the ideal case with $t_2 = t_1$, which assumes that the channel parameters are invariant and will provide BER lower bound for evaluation;  the second scheme adopts the well-used assumption in the OTFS community that $t_2-t_1$ can be regarded as the quasi-stationary interval \cite{OTFS2} ($T_{QS} = 100.8$ ms, calculated based on Section \ref{section:3-1} the CDD (the threshold is greater than 0.9)); the third scheme adopts the quasi-invariant intervals modeled in Section \ref{section:3-3}, corresponding to the minimum quasi-invariant interval $T^{min}_{QI}$ of $5$ main propagation paths under the DD-TCC threshold $\alpha$ = 0.9 (namly, $t_2-t_1 = T^{min}_{QI} = 2.8, 1.87, 1.4$ and $0.93$ ms , respectively). In this analysis, QPSK modulation is employed, the other experimental parameters are same as those in Section \ref{section:5-2}.

The BER performance of schemes with different $t_2-t_1$ is shown in Fig. \ref{fig:BER Comparison across TDDL-A}. First, when comparing the BER of OTFS transmission with $t_1 = t_2$ and $t_2-t_1 = T_{QS}$, we note that the BER will be pretty worse when assuming that the DD domain channel coefficients are invariant during the quasi-stationary interval. Indeed, channel stationarity characterizes the second-order statistics of the channel, from which the invariance of the channel coefficients cannot be directly inferred. Second, for the OTFS transmission with $t_1 - t_2$ set according to the quasi-invariant interval discovered in Section \ref{section:3-3}, the BER shows a trend of gradually approaching the ideal case with the decrease of the minimum quasi-invariant interval, which shows the rationality of the modeled quasi-invariant intervals. Particularly, the BER of OTFS transmission with $t_2-t_1= T^{min}_5$ is closest to that of the ideal case. Accordingly, we can observe that the DD domain channel coefficients are always time-varying even within 0.93 ms. Moreover, we suggest that the DD domain channel can be regarded as invariant when the smallest quasi-invariant interval of multipath is satisfied, which will be on the ms order in the strong time-varying case of the HSR viaduct scenario.

\section{Conclusion}
\label{section:6}
This paper conducts a systematic investigation into DD domain channel measurement and modeling for HSR viaduct scenarios. By leveraging the LTE-R framework, we proposed a novel measurement method enabling commercial OFDM systems to extract DD parameters, and established quasi-stationary channel models across various time-varying conditions. Comprehensive verifications, employing both physical dispersion analysis and system-level BER evaluation, confirm that the proposed models accurately capture the realistic channel characteristics. Crucially, our results demonstrate that the DD domain representation effectively transforms the rapidly time-varying HSR channel into a sparse and quasi-stationary form. This characteristic significantly reduces the complexity of channel estimation while maintaining robust error performance, thereby justifying the preference for DD domain methodologies in next-generation high-mobility communications. In summary, this work provides reliable theoretical and model support for future DDMC system designs.

Based on this work, there are several interesting research directions remain to be explored in the future. First, due to limitations in the measurement setup, we did not investigate the spatial characteristics of the DD domain channel. This aspect could be examined in subsequent studies to enhance the understanding of DD domain channel mechanisms. Second, the current work focuses on DD domain channel modeling in the HSR viaduct scenario, while the characteristics of the DD domain channel in other complex scenarios—such as HSR tunnels, ordinary railways, and urban rail transit—still require further investigation. Moreover, the behavior of the DD domain channel in various integrated sensing and communication (ISAC) scenarios is also of significant interest. Future research based on the proposed designs should be conducted to provide a theoretical foundation and modeling support for the integrated development of ISAC and DDMC technologies in 6G and beyond communication systems.

\bibliographystyle{IEEEtran}
\bibliography{IEEE_TWC_ZH-refs}

\end{document}